\documentclass{article}
\usepackage{graphicx} 
\usepackage{amsmath} 
\usepackage{booktabs} 
\usepackage{subcaption}
\usepackage{float} 
\usepackage{url}
\usepackage{cite}
\usepackage[a4paper, total={6in, 8in}]{geometry}
\usepackage{tikz}
\usetikzlibrary{calc,positioning}

\title{Transforming Japan Real Estate}
\author{Diab Haque}
\date{April 2024}

\begin{document}

\maketitle

\begin{abstract}
The Japanese commercial and residential real estate markets, with a combined estimated value surpassing 35 trillion US dollars, present a significant opportunity for investors. Accurate forecasting of rents and prices could provide a substantial competitive edge in this market. This paper investigates the potential of using alternative data variables to predict the performance of real estate in 1100 municipalities across Japan.

To test this hypothesis, a comprehensive house price index is constructed for all Japanese municipalities from 2005 to the present, utilizing a dataset containing over 5 million transactions. This core dataset is then enriched with economic factors for each municipality spanning the past few decades. Using this enhanced dataset, price trajectories are predicted for each municipality.

The findings indicate that alternative data variables can indeed forecast real estate performance. Investment signals based on these variables yielded considerable positive returns with low volatility. For instance, the net migration ratio delivered an annualized return of 4.6\% with a Sharpe ratio of 1.5, while taxable income growth and new dwellings ratio recorded annualized returns of 4.1\% (Sharpe ratio of 1.3) and 3.3\% (Sharpe ratio of 0.9), respectively. Furthermore, when these variables are combined with the power of transformer models to predict risk-adjusted returns 4 years in advance, the model achieves an R-squared score of 0.28, explaining nearly 30 percent of the variation in future municipality prices.

These findings underscore the need for further research to identify additional factors that can enhance the prediction of real estate performance. Nevertheless, the results provide compelling evidence that alternative data variables can be valuable tools for real estate investors. By harnessing these data sources, investors can gain deeper insights into the drivers of real estate prices and make more informed investment decisions in the Japanese market. 

\end{abstract}

\newpage

\section{Introduction} \label{intro}
The Japanese commercial and residential real estate markets, valued at over 35 trillion US dollars, attract a diverse range of participants. While investors employ various strategies to generate returns, profits ultimately stem from future rents and disposition prices, adjusted for the time value of money, net of the current acquisition price. Mathematically, this can be expressed as:
\begin{equation}
    \text{Net Profit} = \sum_{i = 1}^{n - 1} \frac{\text{rent}_i}{(1 + r)^i} + \frac{\text{price}_n}{(1 + r)^n} - \text{price}_0
\end{equation}

Accurate forecasting of rents and prices would provide a significant competitive advantage to investors, assuming their competitors lack access to the same information.

Meszaros (2024) \cite{Meszaros2024} emphasizes that real estate dynamics operate more distinctly at a local rather than a national level—a view supported by anecdotal evidence from Asaftei et. al. (2018) \cite{asaftei2018bigdata}, who noted that properties in proximity to Starbucks in Boston significantly outperformed other local markets. This phenomenon underscores the importance of micro-location in property valuation, an area that remains underexplored in Japanese real estate literature.

Although extensive research, such as that by Chen et al. (2013) \cite{chen2013moody}, exists on forecasting house price indices in the US using economic, demographic, and historical data, Japan presents a gap in localized, data-driven forecasting research. Shimizu's studies \cite{Shimizu2010, Shimizu2007} on house price indices primarily concentrates on a specific subset of locations like the 23 wards of Tokyo, offering limited insights into Japan’s broader municipalities. Moreover, while Yoshihiro et al. (2017) \cite{Yoshihiro2017} explored demographic impacts on housing prices on a wider array of municipalities, their simulation-based approach was oriented towards long-term demographic trends over decades rather than actionable investment insights.

This paper aims to bridge this gap by testing the predictive power of alternative data variables on real estate performance at a micro-location level in Japan. We hypothesize that these alternative data sets will provide more granular insights into market dynamics, thereby offering a competitive edge to investors. The structure of the remaining paper is as follows:

Chapter 2 defines real estate performance. Real estate is a heterogeneous asset class and price determination is not like that of standardized securities. We employ transaction datasets to quantify market performance, illustrating trends such as the doubling of real estate prices in Tokyo's Minato-shi or the 30\% decline in Chiba-ken’s Isumi city over the past decade.

Chapter 3 investigates some potential drivers of property prices, detailing the data sources, necessary processing, and correlation with our price metrics. We also devise simple long-short strategies to demonstrate the alpha in these factors.

Chapter 4 explores the application of machine learning to time series analysis. It utilizes Transformer models to predict derived prices using a temporal window of factors, introduces a framework for training and testing on time series datasets, and analyses model results to demonstrate performance.

Chapter 5 concludes. 

\newpage

\section{Price Determination} \label{price}

Real estate is a heterogeneous asset class, with no two properties being identical. As the distance between properties increases, they tend to exhibit increasingly divergent price patterns. However, there is a general consensus that real estate submarkets have a notion of performance. This chapter aims to capture this performance through a price index for each submarket, calculated using transaction data provided by the Ministry of Land, Infrastructure, Transport and Tourism (MLIT).

\subsection{Dataset}

The dataset spans from Q4 2005 to the present and contains over 5 million transactions across all 47 prefectures in Japan. Each transaction is represented by a row with 30 columns describing various attributes of the transaction. The data is collected through surveys conducted by the MLIT \cite{mlit_nd}.

\subsection{Location}

To define real estate markets or locations, we use municipalities as our geographic divisions. Japanese municipalities are local administrative units, typically averaging around 170 square kilometres in size. Our dataset covers approximately 1700 municipalities, of which about 1100 have sufficient transactions for analysis. The choice of municipalities as the geographical unit is apt. If it is too large, say to the prefecture level, we lose a lot of information on aggregation and the variations in prices of municipalities tend to cancel out - causing the index fluctuations to be muted. If it is too small, we significantly reduce the factor data available to us (for example, the datasets from the SSDS described in chapter \ref{factors} are only granular down to the municipality level), and the number of transactions may become too small for price indices to be reliable.

\subsection{Price Index Methodology}

We employ a quality-adjusted price index using the time dummy variable method (also known as the hedonic regression method). This method estimates the relationship between property prices and various factors such as size, location, age, and usage. By doing so, it isolates pure price changes from changes in the composition and quality of properties sold during a given period. Each time period is assigned a weight corresponding to its influence on prices, which forms our price index. This method is close to that used by the MLIT \cite{jrpippi2020} to compose the Japan Residential Property Price Index (JRPPI). The difference is that this study does not compute a rolling monthly index whereas the JRPPI does.

\subsubsection{Data Pre-processing}

\begin{figure}
    \centering
    \includegraphics[width=0.9\linewidth]{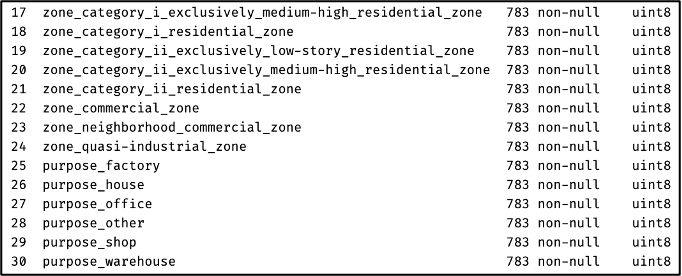}
    \caption{A subset of the categorical columns for transactions data}
    \label{fig:categorical_column_subset}
\end{figure}

The dataset contains both numerical and categorical factors, with many fields being null or containing multiple categorical values in one cell. To handle this, we pre-process each field, resulting in up to 75 columns for a single area code. To reduce dimensionality and address potential multicollinearity issues, we perform Principal Component Analysis (PCA) on the factors, reducing the data-frame to a half or a third of its original size while retaining 95\% of the variance. Finally, we concatenate the principal components with time dummy variables for each year, preparing the dataset for regression. Figure~\ref{fig:categorical_column_subset} shows a subset of the categorical columns and illustrates why the dataset becomes sparse.

\subsubsection{Regression Model}

The dependent variable in our model is the natural logarithm of the trade price per unit area, while the independent variables include the principal components and time dummy variables. The model is represented as follows:

\begin{equation}
    \ln(p) = \beta_0 + \sum_{i=1}^t \gamma_i D_i + \sum_{j=1}^n \beta_j PC_j + \varepsilon
\end{equation}

Where:
\begin{itemize}
  \item $p$: price per unit area for the transaction
  \item $\beta_0$: constant term
  \item $\gamma_i$: time dummy parameter at year $i$
  \item $D_i$: time dummy variable (1 at the year of transaction and 0 otherwise)
  \item $\beta_j$: parameter for principal component $j$
  \item $PC_j$: principal component $j$
  \item $\varepsilon$: error term
\end{itemize}

\begin{figure}
    \centering
    \includegraphics[width=0.9\linewidth]{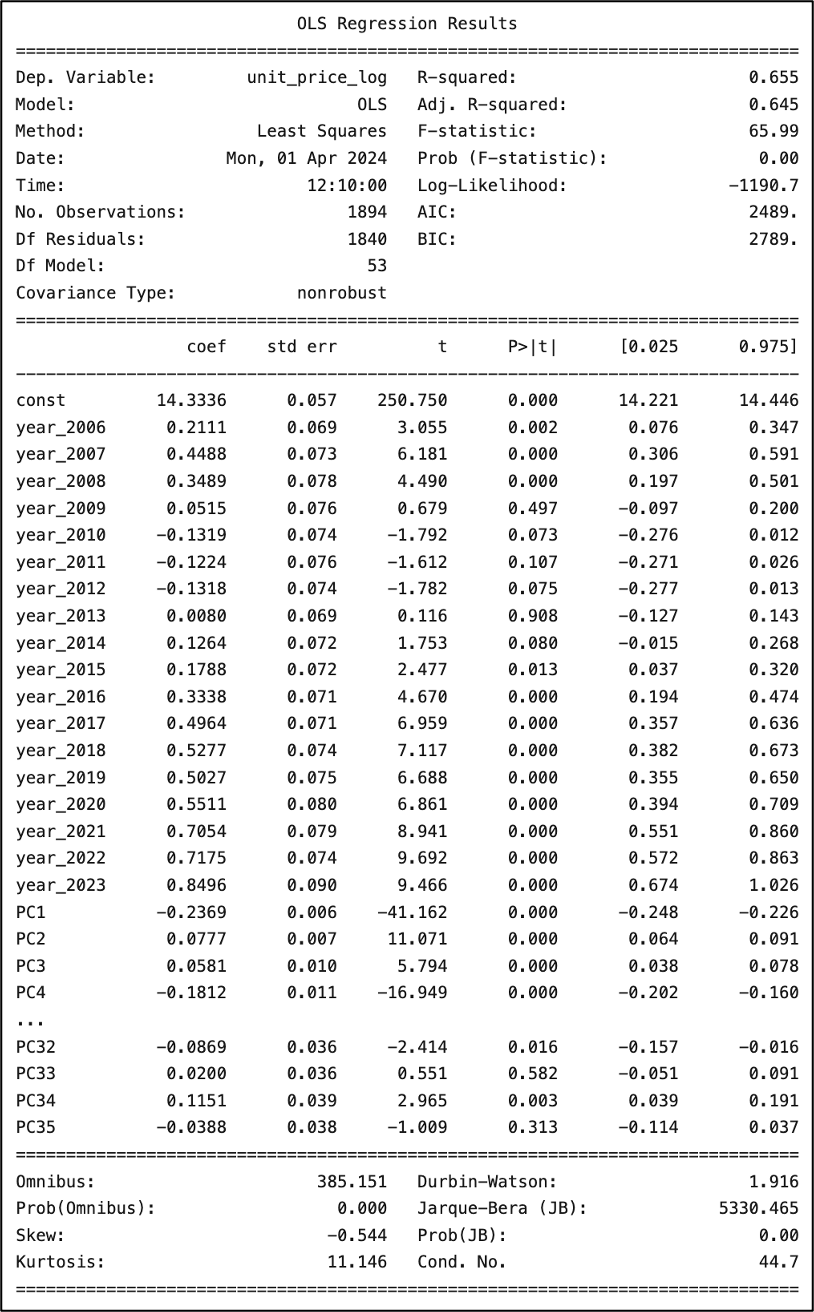}
    \caption{Results of OLS regression}
    \label{fig:ols_results}
\end{figure}

Figure~\ref{fig:ols_results} shows the results of an ordinary least squares regression run on our transactions dataset. This produces the coefficients for our year dummies and p-values that represent the confidence for these coefficients.

\subsubsection{Price Index Conversion}

The coefficients of the time dummy variables are exponentiated and divided by the value of the base year (the second available year for the municipality) to create the price index:

\begin{equation}
    \text{Price Index}_t = \dfrac{\exp{\gamma_t}}{\exp{\gamma_1}}
\end{equation}

\begin{figure}
    \centering
    \includegraphics[width=0.9\linewidth]{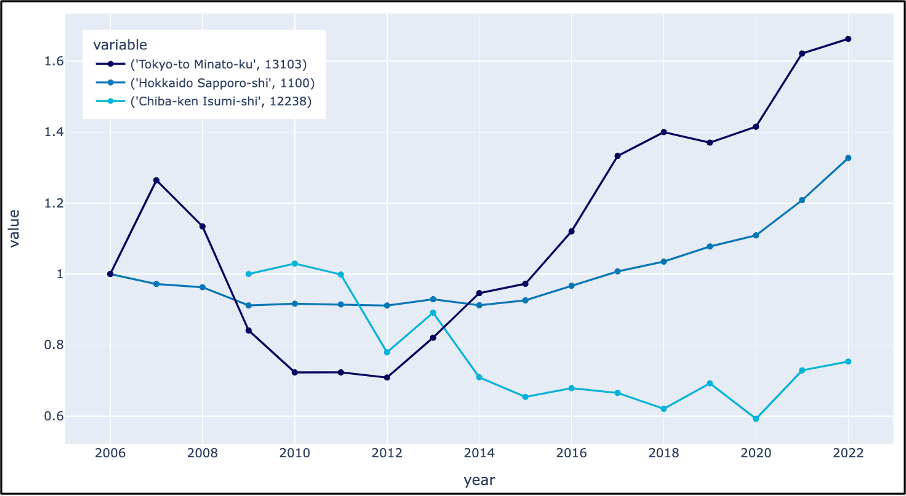}
    \caption{Residential Price index for selected municipalities over time}
    \label{fig:price_indices}
\end{figure}

Table~\ref{tab:price_index_data} shows the resulting time series dataset for a single municipality. This is compiled for every municipality, thereby creating a cross-sectional time series dataset (also known as a panel dataset). Figure~\ref{fig:price_indices} illustrates the price indices for 3 such municipalities over time. The chart clearly demonstrates the varying price patterns displayed by these municipalities.

\begin{table}[h]
\centering
\begin{tabular}{ccccc}
\toprule
Year & Area Code & Area & Price Index & YoY Growth \\
\midrule
2006 & 13103 & Tokyo-to Minato-ku & 100.00 & \\
2007 & 13103 & Tokyo-to Minato-ku & 126.84 & 0.27 \\
2008 & 13103 & Tokyo-to Minato-ku & 114.78 & -0.10 \\
2009 & 13103 & Tokyo-to Minato-ku & 85.25 & -0.26 \\
2010 & 13103 & Tokyo-to Minato-ku & 70.97 & -0.17 \\
2011 & 13103 & Tokyo-to Minato-ku & 71.64 & 0.01 \\
\ldots & \ldots & \ldots & \ldots & \ldots \\
2018 & 13103 & Tokyo-to Minato-ku & 137.24 & 0.03 \\
2019 & 13103 & Tokyo-to Minato-ku & 133.86 & -0.02 \\
2020 & 13103 & Tokyo-to Minato-ku & 140.49 & 0.05 \\
2021 & 13103 & Tokyo-to Minato-ku & 163.94 & 0.17 \\
2022 & 13103 & Tokyo-to Minato-ku & 165.93 & 0.01 \\
2023 & 13103 & Tokyo-to Minato-ku & 189.37 & 0.14 \\
\bottomrule
\end{tabular}
\caption{Price index for Tokyo-to Minato-ku}
\label{tab:price_index_data}
\end{table}

\subsubsection{Assumptions}

The time dummy variable method has advantages, such as controlling for property quality and heterogeneity. However, it assumes that the relationship between property characteristics and prices remains constant over time and requires a large, representative sample of transactions to produce reliable estimates.

While the computed price indices serve as a foundation for our research, refining the methodology could yield substantial enhancements. Improving the precision and accuracy of these indices would provide a more robust and reliable basis for drawing insights and conclusions.

In the next chapter, we will use these "prices" to analyse variations in growth quantitatively and explore economic factors that may explain them.

\newpage

\section{Factors} \label{factors}

The determination of real estate prices is influenced by the interplay of factors reflecting the conditions of supply and demand within specific markets. We aim to identify and quantify these determinants, and use them to predict real estate valuation of a submarket.

There exist a myriad of variables that can encapsulate both supply and demand. On the supply side, there are variables such as the number of buildings for sale and under construction. On the demand side, there is population growth, number of new businesses and enterprises, job creation, education levels, crime rates, traffic congestion, neighbourhood walkability, market sentiment and more.

The factors we choose must be quantifiable, and to test them, we must find their values over time for each location. Table~\ref{tab:panel_data_form} shows the schema both our price and factor datasets must conform to, i.e. they must be cross-sectional time series datasets indexed by area code and year.

\begin{table}[ht]
\centering
\begin{tabular}{ccc}
\toprule
Year & Area Code & Value \\
\midrule
\ldots & \ldots & \ldots \\
\ldots & \ldots & \ldots \\
\bottomrule
\end{tabular}
\caption{Required form for panel dataset}
\label{tab:panel_data_form}
\end{table}

The System of Social and Demographic Statistics (SSDS) created by the Statistics Bureau of Japan offers data in this format, although it does not cover all the factors we would like to explore. Nevertheless, this platform serves as a good starting point in the potentially endless search for alternative datasets.

Each section in this chapter starts by introducing a factor individually - discussing prior research done on the affect of the factor on house prices. We discover that this dive into prior research often uncovers contradicting research outcomes, and it becomes apparent why it is so easy to commit the sin of storytelling (Wang et. al., 2014 \cite{Wang2014}). While we can surely draw insights from these results, the reader should remember that it is easy to explain most statistical results after the fact. 

The section then explains the collection and pre-processing of the dataset from its sources and visualizes them for the reader.

Next, the analysis constructs simple linear models to demonstrate the influence of each factor on returns succinctly. The linear model utilizes the cumulative growth of the factor over the three years preceding any given time period \( t \), to predict the returns \( k \) years ahead at time period \( t+k \). This approach enables testing of hypotheses such as, "The population in our municipality has increased by 20\% over the last \( N \) years; I predict housing prices will rise in the next \( n \) years." Although this step enriches the analysis, it also highlights the limitations of single-factor linear models in predicting returns, as evidenced by low $R^2$ scores which indicate significant noise in the dependent variable.

To conclude the section, we evaluate a series of simple investment strategies based on individual factors. We initiate long positions in municipalities exhibiting the highest cumulative growth for a given factor and short positions in those with the lowest growth. Annually, we re-mark our positions using the house price index and hold (rebalance) these positions over \( n \) years, where \( n \) remains constant for each factor within an experiment but varies between experiments. Subsequently, we plot the return (growth in NAV) over time of our hypothetical factor-based portfolio against the returns of a thousand portfolios constructed using random noise as an investment signal. The investment universe is restricted to the 500 most populous municipalities in Japan, with all conditions being identical across different factors. The objective is to discern potential relationships between our selected factors and returns, which cannot be attributed merely to random chance.

Please note that the methods employed are not formal hypothesis tests. While several statistical tests for cross-sectional time series data, such as the panel data Granger causality tests developed by Dumitrescu and Hurlin (2012) \cite{Dumitrescu2012}, were explored, their use did not necessarily improve confidence in the hypotheses or understanding of the interplay between factors and returns.

\subsection{Population and Migration}

There is a large body of literature that investigates the effect of internal migration on house prices across various countries. The results vary and the explanations are nuanced. Saiz (2006) \cite{Saiz2006} argues that immigration pushes up the demand for housing, promoting an increase in rents and subsequently, an increase in housing prices. Concretely, his research on American cities suggests that a 1\% increase in immigration results in roughly a 1\% growth in home prices for American cities. Filipa Sá’s (2011) \cite{Sa2011} studies contrasts this. Sa extended Saiz’s model to include the effect of income on housing demand and the preference of natives for immigration. She claimed that the downward pressure from an outflow of natives (due to increased migration) would counterbalance and possibly overwhelm the upward pressure on prices caused by the inflow of immigrants and other natives. Her study showed a negative correlation between prices and immigration in the UK. Several such studies have been conducted across the globe, and results fall into a spectrum between significant negative correlation to significant positive correlation.

\subsubsection{Dataset}

The population and migration datasets are obtained from the SSDS. They start at 1996 and end at 2022, with yearly values. The population dataset is reported from the national population census, while the migration dataset is composed from the Basic Resident Registration which is prepared in accordance with the Basic Resident Registration Act. It is reported by the Statistics Bureau of the Ministry of Internal Affairs and Communications of Japan \cite{statsbureau_migration}.

Once the dataset is cleaned, we can get the net of the migrations in and out of an area code, and divide that by the total population to get the net migration ratio.

\begin{equation}
    Net\ Migration\ Ratio = \frac{In\ Migrations - Out\ Migrations}{Total\ Population}
\end{equation}

\begin{figure}
    \centering
    \includegraphics[width=0.90\linewidth]{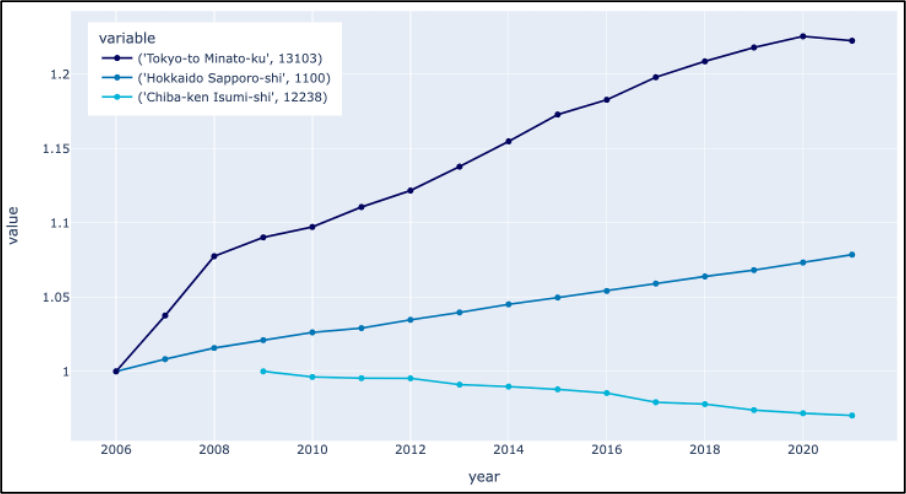}
    \caption{Cumulative net migration for selected municipalities}
    \label{fig:cum_net_migration}
\end{figure}

Figure~\ref{fig:cum_net_migration} illustrates the migration patterns for our original three municipalities.

\subsubsection{Simple Linear Model}

\begin{figure} 
  \begin{subfigure}[b]{0.5\linewidth}
    \centering
    \frame{\includegraphics[width=0.95\linewidth]{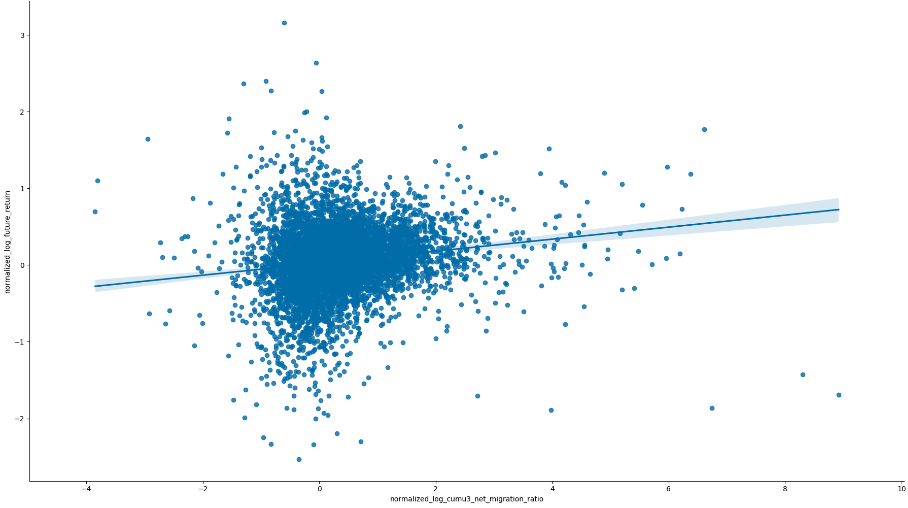}}
    \label{fig:net_migration_ols:a} 
    \vspace{2ex}
  \end{subfigure}
  \begin{subfigure}[b]{0.5\linewidth}
    \centering
    \frame{\includegraphics[width=0.95\linewidth]{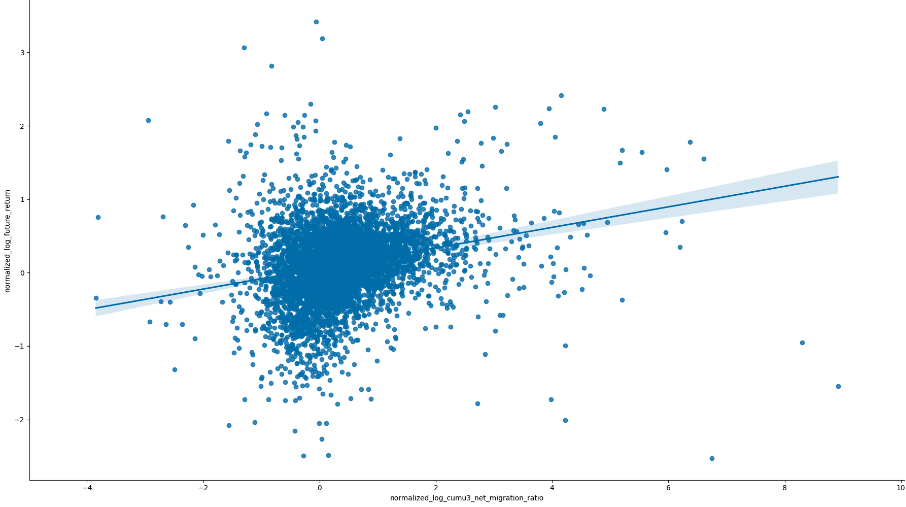}} 
    \label{fig:net_migration_ols:b} 
    \vspace{2ex}
  \end{subfigure} 
  \begin{subfigure}[b]{0.5\linewidth}
    \centering
    \frame{\includegraphics[width=0.95\linewidth]{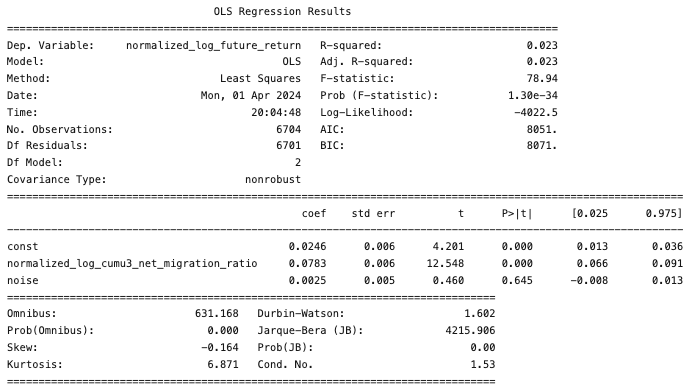}} 
    \caption{\footnotesize \textbf{Horizon: 2y} – $R^2$: 0.006; Coef: 0.0469;} 
    \label{fig:net_migration_ols:c} 
  \end{subfigure}
  \begin{subfigure}[b]{0.5\linewidth}
    \centering
    \frame{\includegraphics[width=0.95\linewidth]{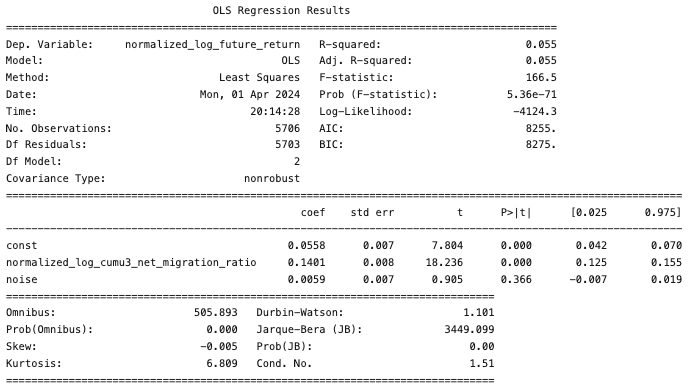}} 
    \caption{\footnotesize \textbf{Horizon: 4y} – $R^2$: 0.016; Coef: 0.8853;} 
    \label{fig:net_migration_ols:d} 
  \end{subfigure} 
  \caption{Results of linear model for taxable income growth}
  \label{fig:net_migration_ols} 
\end{figure}

Figure~\ref{fig:net_migration_ols} suggests a positive correlation between the cumulative migration over the trailing 3 years and the returns of the municipality. This positive correlation increases in magnitude and explanatory power as the time horizons increases from 2 years to 4. The regression results show that the p-values of the coefficients are significant, and so is the F-score for the model.

\subsubsection{Long Short Strategy}

\begin{figure} 
  \begin{subfigure}[b]{0.5\linewidth}
    \centering
    \frame{\includegraphics[width=0.95\linewidth]{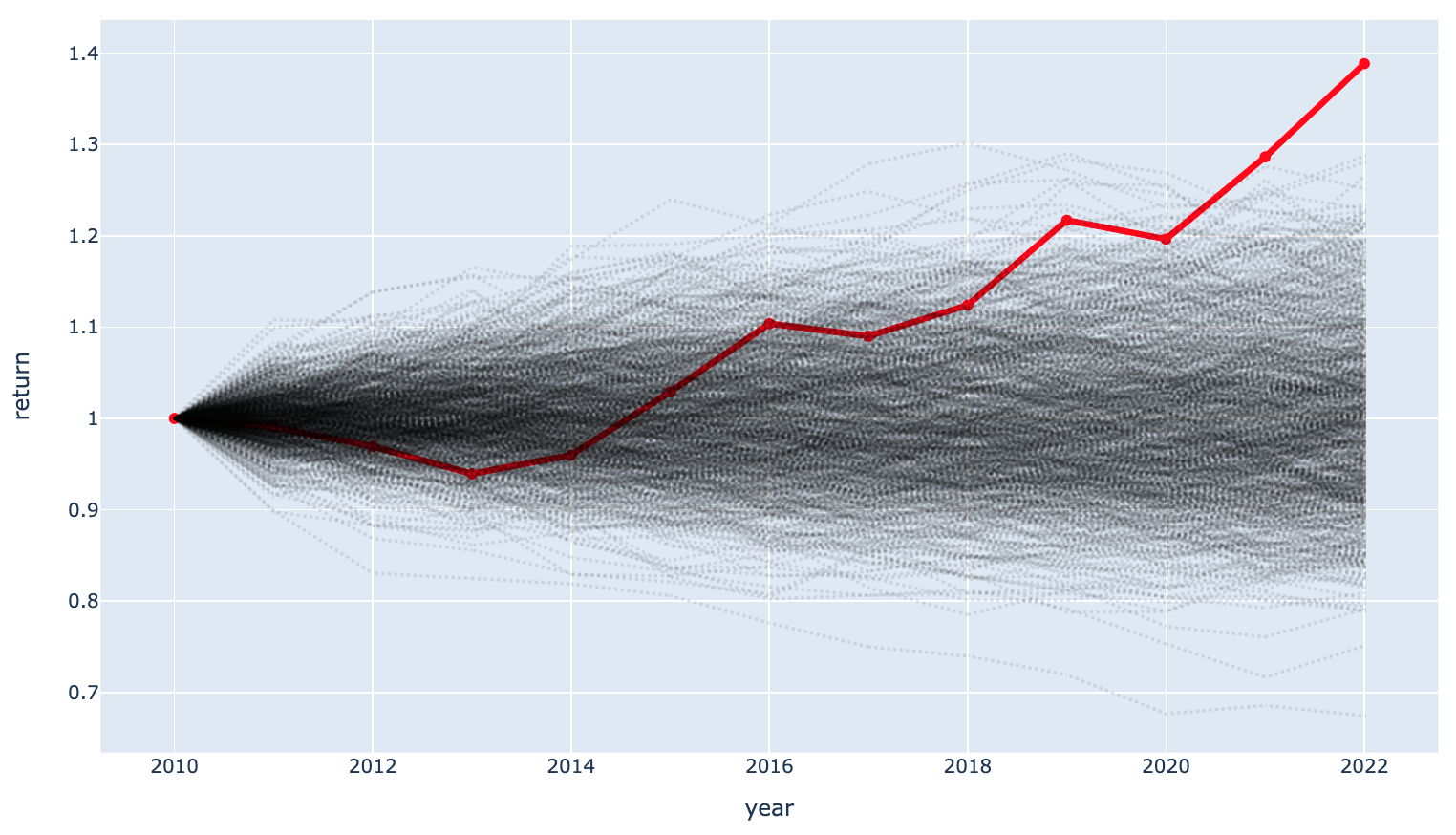}}
    \label{fig:net_migration_long_short:a} 
    \caption{\footnotesize \textbf{Horizon: 1y} - CAGR: 2.56\%; Sharpe: 0.66;} 
    \vspace{2ex}
  \end{subfigure}
  \begin{subfigure}[b]{0.5\linewidth}
    \centering
    \frame{\includegraphics[width=0.95\linewidth]{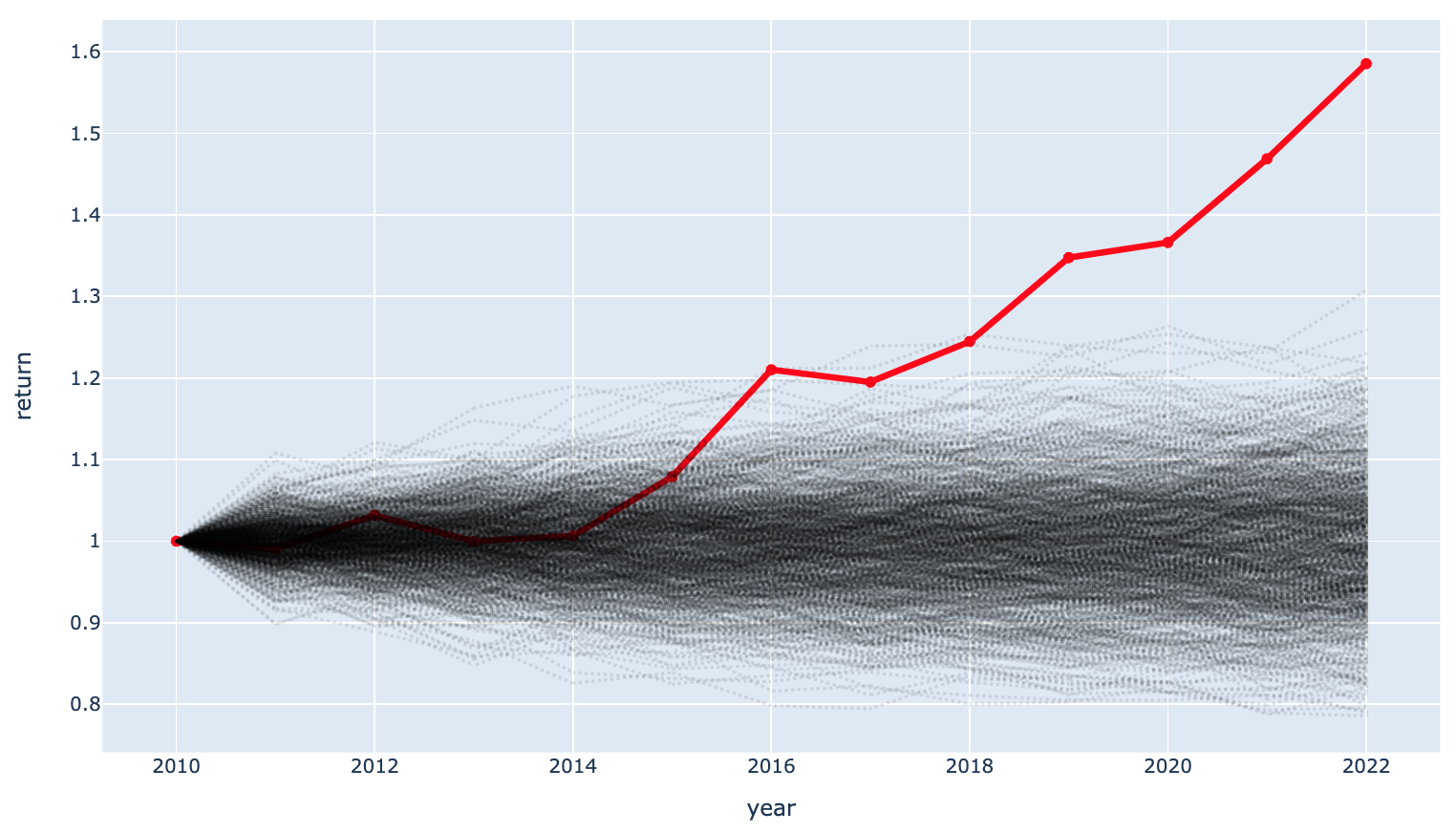}} 
    \label{fig:net_migration_long_short:b} 
    \caption{\footnotesize \textbf{Horizon: 2y} - CAGR: 3.61\%; Sharpe: 0.89;} 
    \vspace{2ex}
  \end{subfigure} 
  \begin{subfigure}[b]{0.5\linewidth}
    \centering
    \frame{\includegraphics[width=0.95\linewidth]{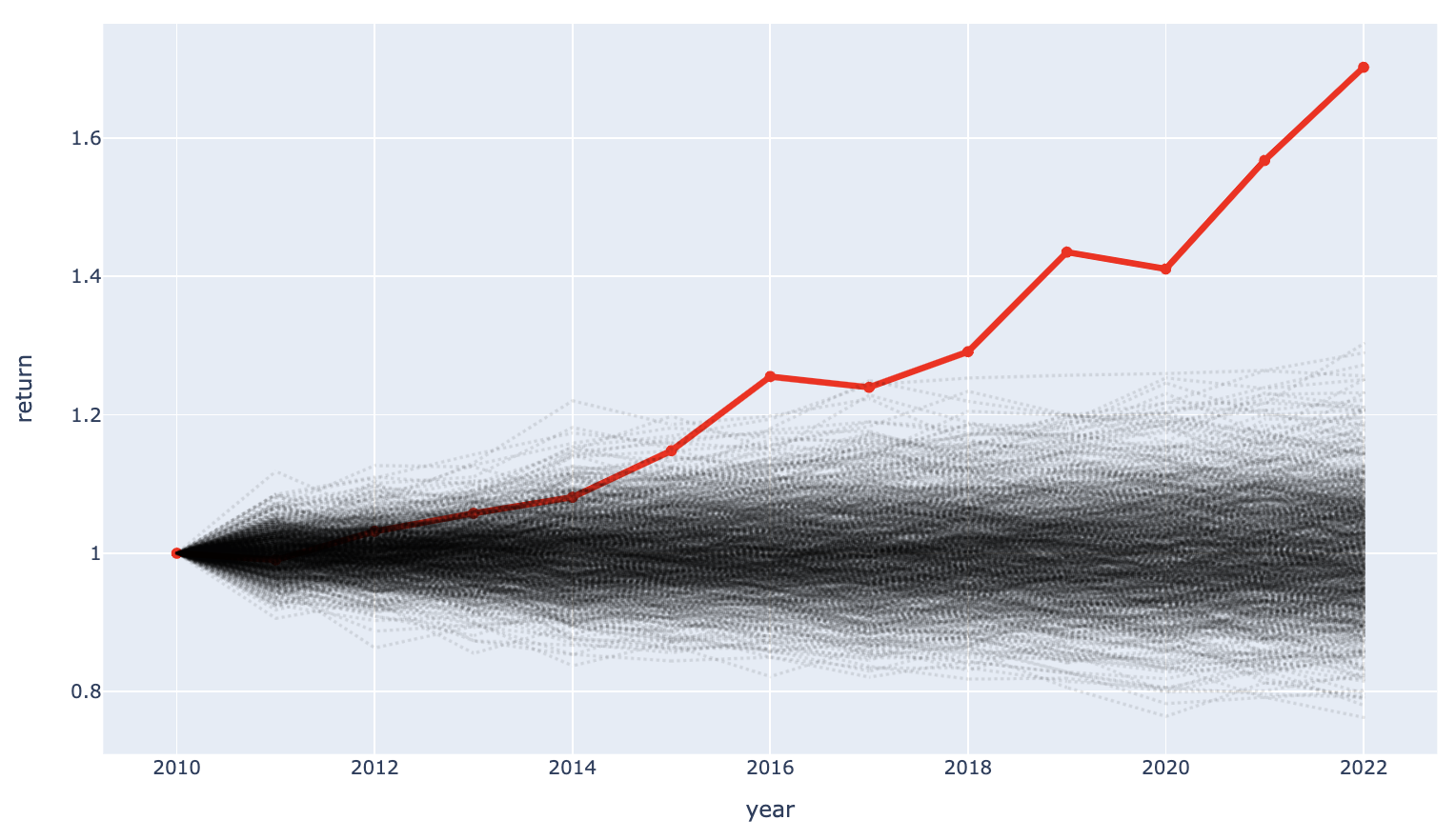}} 
    \caption{\footnotesize \textbf{Horizon: 3y} - CAGR: 4.18\%; Sharpe: 1.03;} 
    \label{fig:net_migration_long_short:c} 
  \end{subfigure}
  \begin{subfigure}[b]{0.5\linewidth}
    \centering
    \frame{\includegraphics[width=0.95\linewidth]{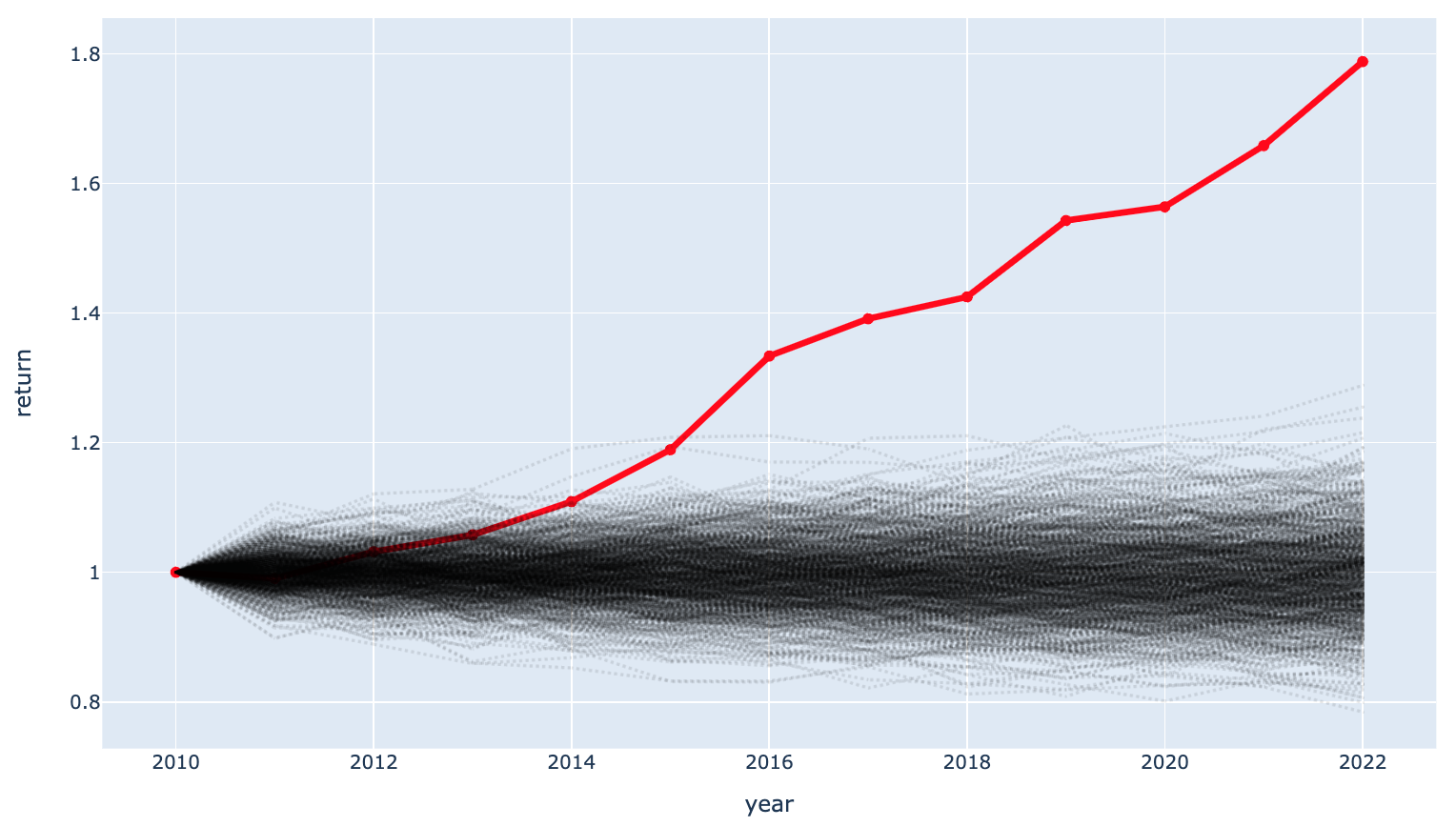}} 
    \caption{\footnotesize \textbf{Horizon: 4y} - CAGR: 4.57\%; Sharpe: 1.48;} 
    \label{fig:net_migration_long_short:d} 
  \end{subfigure} 
  \caption{Returns from net migration ratio as a signal}
  \label{fig:net_migration_long_short} 
\end{figure}

Figure~\ref{fig:net_migration_long_short} shows the performance of our variable over holding periods ranging from 1 to 4 years. We notice that net migration serves as a strong predictor of growth in Japan over the longer term. While external factors make it difficult to predict with certainty whether price of an individual location will move up with greater net migrations, we see that regions with higher net migrations demonstrate significantly greater returns on average.

\newpage
\subsection{Taxable Income}

Similar to migrations, research has been conducted for decades on the relationship between income and house prices - and once again, the conclusions are not simply interpretable. While many housing market observers believe that house prices and income are cointegrated and the two variables return to a long-term equilibrium, Gallin (2006) \cite{Gallin2006} asserts that such assumptions are unjustified. Using city level data over 23 years and the results of a set of standard tests, he states that the data does not display evidence of cointegration between the two variables. That said, he does not dismiss the possibility of a relationship between the two variables, saying rather that we cannot verify such a claim with the data available.

\subsubsection{Dataset}

The Taxable income dataset starts at 1985 and ends at 2021. It includes both the number of taxpayers and the total taxable income. It is collected by the Local Tax Bureau and provided by the Statistics Bureau through the SSDS.

\begin{figure}
    \centering
    \includegraphics[width=0.90\linewidth]{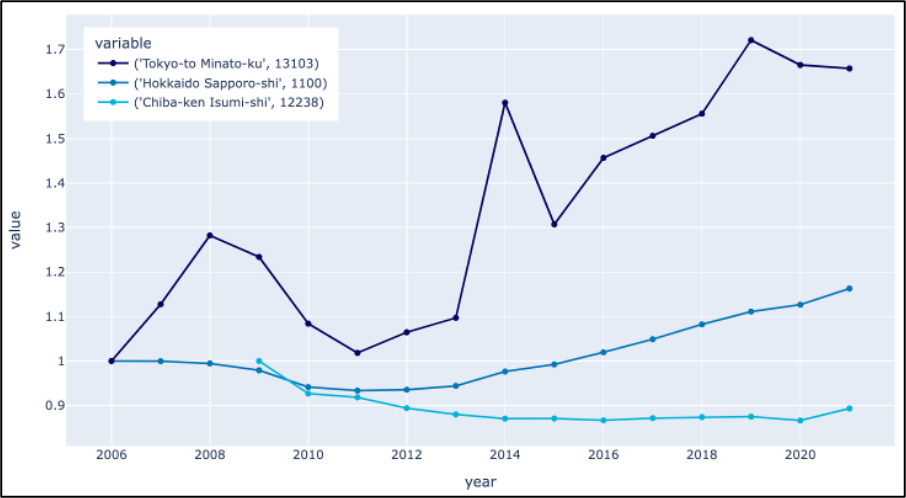}
    \caption{Cumulative taxable income growth for selected municipalities}
    \label{fig:cum_taxable_income}
\end{figure}

With the dataset cleaned, we find the growth by taking the percentage change over each period. Figure~\ref{fig:cum_taxable_income} visualizes how taxable income income for our selected municipalities varies over time.

\subsubsection{Simple Linear Model}

\begin{figure} 
  \begin{subfigure}[b]{0.5\linewidth}
    \centering
    \frame{\includegraphics[width=0.95\linewidth]{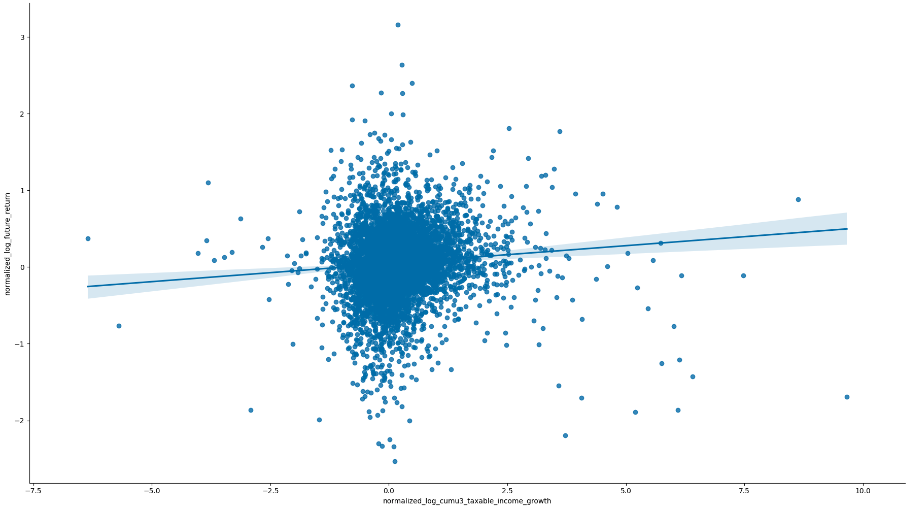}}
    \label{fig:taxable_income_ols:a} 
    \vspace{2ex}
  \end{subfigure}
  \begin{subfigure}[b]{0.5\linewidth}
    \centering
    \frame{\includegraphics[width=0.95\linewidth]{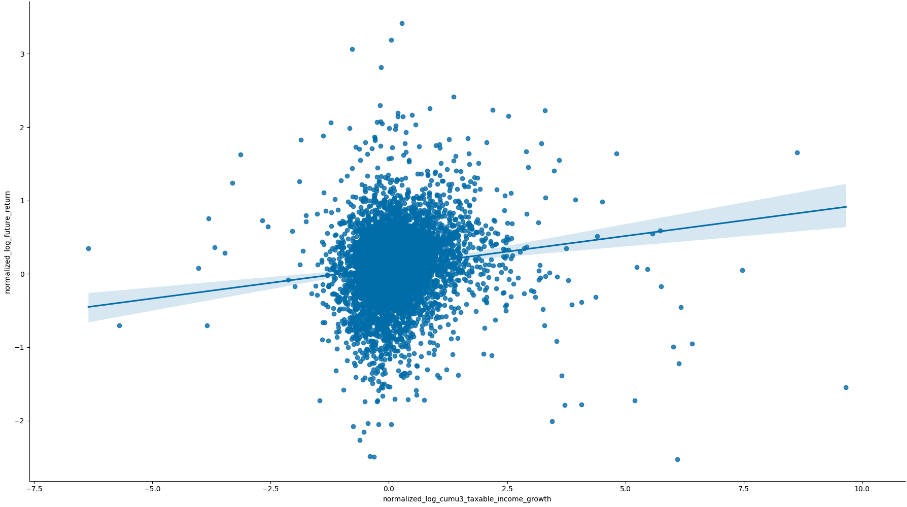}} 
    \label{fig:taxable_income_ols:b} 
    \vspace{2ex}
  \end{subfigure} 
  \begin{subfigure}[b]{0.5\linewidth}
    \centering
    \frame{\includegraphics[width=0.95\linewidth]{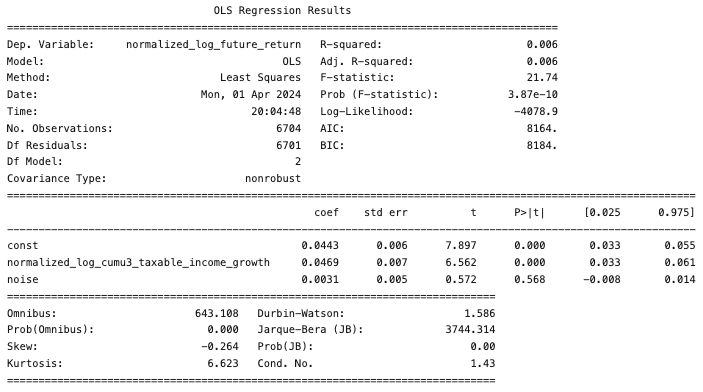}} 
    \caption{\footnotesize \textbf{Horizon: 2y} – $R^2$: 0.023; Coef: 0.0783;} 
    \label{fig:taxable_income_ols:c} 
  \end{subfigure}
  \begin{subfigure}[b]{0.5\linewidth}
    \centering
    \frame{\includegraphics[width=0.95\linewidth]{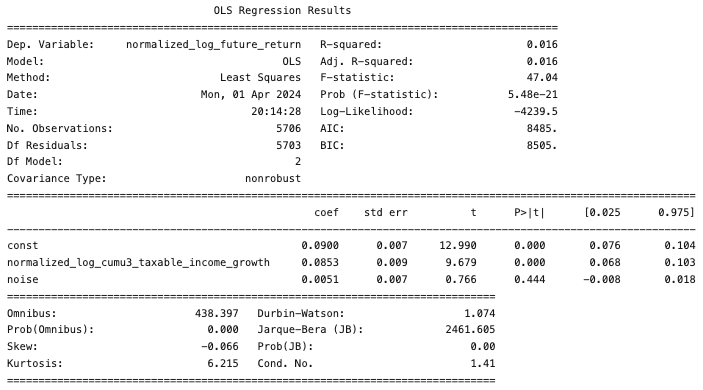}} 
    \caption{\footnotesize \textbf{Horizon: 4y} – $R^2$: 0.055; Coef: 0.1401;} 
    \label{fig:taxable_income_ols:d} 
  \end{subfigure} 
  \caption{Results of linear model for taxable income growth}
  \label{fig:taxable_income_ols} 
\end{figure}

The results of the linear model for taxable income (demonstrated in Figure~\ref{fig:taxable_income_ols}) are significant. They also suggests a positive correlation, and the ability to signal returns do also increase over an increasing time horizon.

\subsubsection{Long Short Strategy}

\begin{figure} 
  \begin{subfigure}[b]{0.5\linewidth}
    \centering
    \frame{\includegraphics[width=0.95\linewidth]{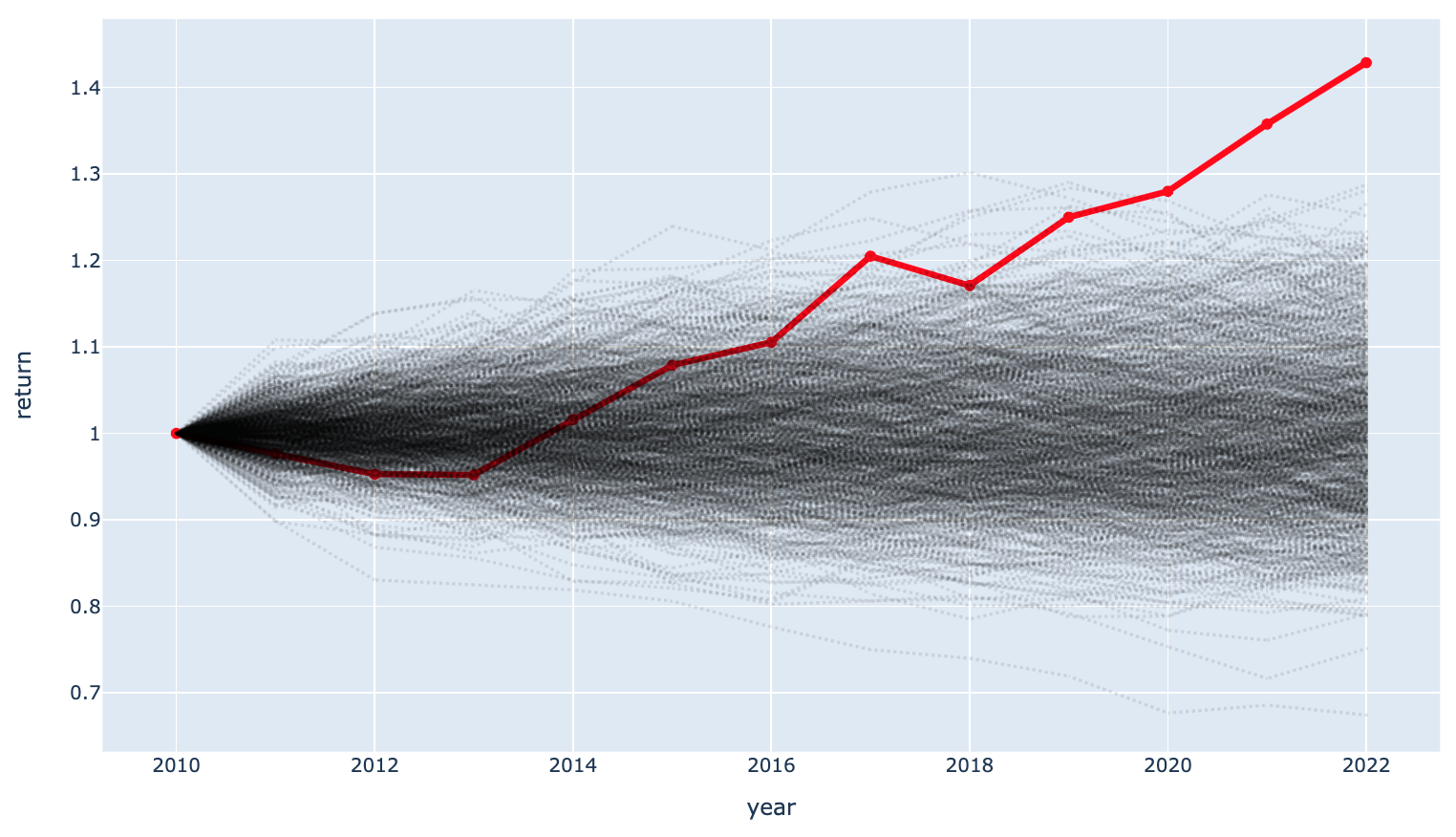}}
    \label{fig:taxable_income_long_short:a} 
    \caption{\footnotesize \textbf{Horizon: 1y} - CAGR: 1.78\%; Sharpe: 0.78;} 
    \vspace{2ex}
  \end{subfigure}
  \begin{subfigure}[b]{0.5\linewidth}
    \centering
    \frame{\includegraphics[width=0.95\linewidth]{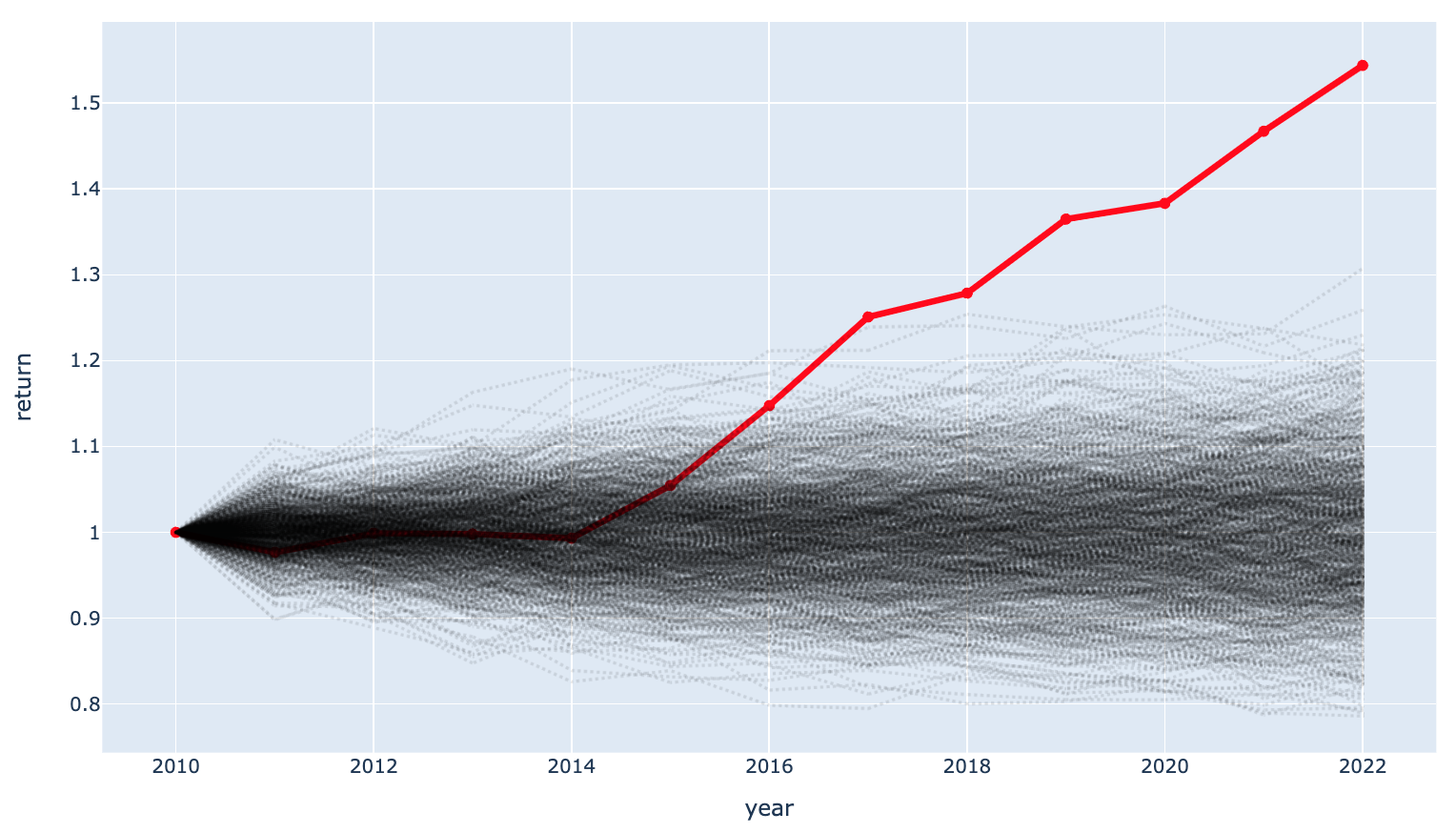}} 
    \label{fig:taxable_income_long_short:b} 
    \caption{\footnotesize \textbf{Horizon: 2y} - CAGR: 3.40\%; Sharpe: 1.04;}
    \vspace{2ex}
  \end{subfigure} 
  \begin{subfigure}[b]{0.5\linewidth}
    \centering
    \frame{\includegraphics[width=0.95\linewidth]{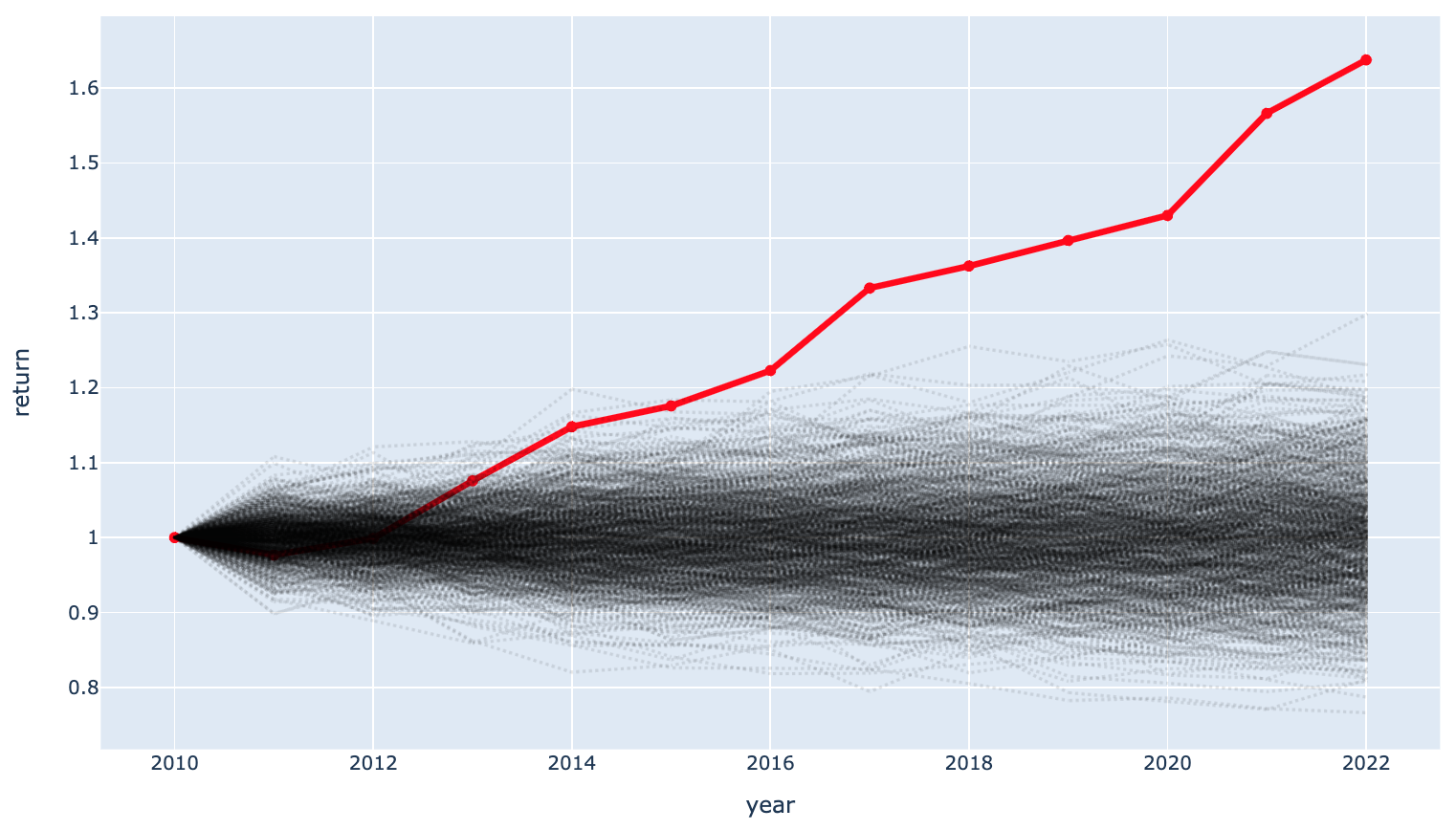}} 
    \caption{\footnotesize \textbf{Horizon: 3y} - CAGR: 3.87\%; Sharpe: 1.29;} 
    \label{fig:taxable_income_long_short:c} 
  \end{subfigure}
  \begin{subfigure}[b]{0.5\linewidth}
    \centering
    \frame{\includegraphics[width=0.95\linewidth]{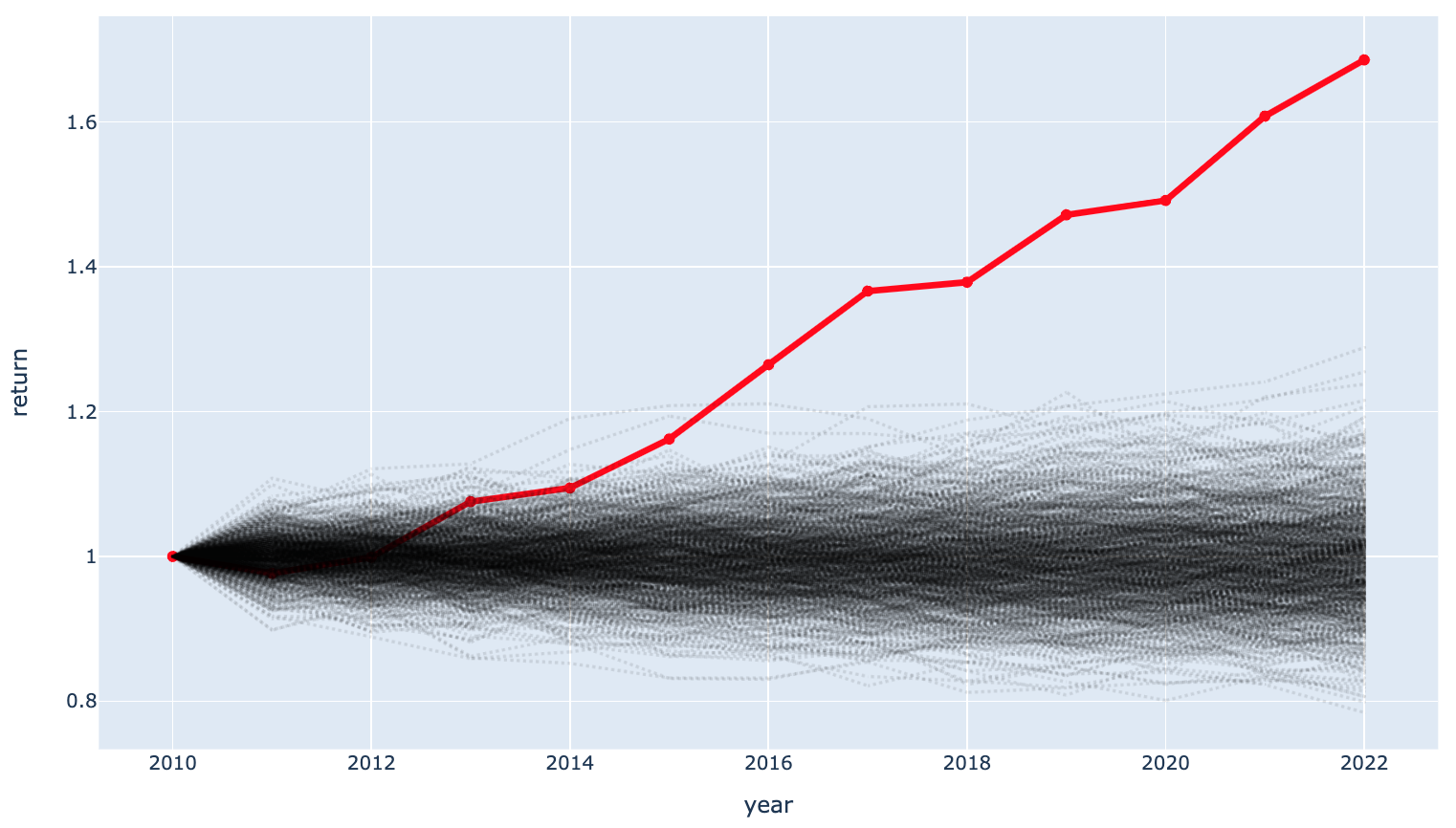}} 
    \caption{\footnotesize \textbf{Horizon: 4y} - CAGR: 4.10\%; Sharpe: 1.31;} 
    \label{fig:taxable_income_long_short:d} 
  \end{subfigure}

  \caption{Returns from taxable income growth as a signal}
  \label{fig:taxable_income_long_short} 
\end{figure}

Similar to net migrations, our fundamental signal of cumulative taxable income growth shows a clear ability to predict returns over the long term (Figure~\ref{fig:taxable_income_long_short}).

We should note that taxable income is an incredibly broad demographic indicator. I believe it would be quite insightful to break taxable income down to its constituent components and use those more granular factors for our analysis. That may provide us an improved theoretical framework of what actually which is driving real estate returns.

\newpage
\subsection{Newly Constructed Dwellings}

The number of newly constructed dwellings is a partial signal of supply into the housing market. Unlike net migrations and taxable income, which primarily affect the demand side of the equation (albeit with potential indirect effects on supply), the construction of new dwellings directly contributes to the available housing stock. The reason it is partial is because the total supply of houses in a market is determined by the sum of existing houses up for sale and newly built units.

According to the fundamental principles of supply and demand, an increase in the number of new dwellings should exert downward pressure on housing prices, assuming constant demand. Moreover, research by Glaeser et al. (2008) \cite{Glaeser2008} suggests that the elasticity of housing supply plays a crucial role in the duration of housing bubbles. Markets characterised by a more elastic housing supply, where new construction can quickly respond to increased demand, tend to experience shorter periods of price inflation compared to markets with inelastic supply.

\subsubsection{Dataset}

The New Dwellings dataset consist of two portions. One is the housing and land survey which is conducted at five year intervals by the Statistic Bureau and data is available from 1998 to 2018. This provides us the number of existing dwellings in a certain municipality. The other is the number of new construction starts which is published annually by the MLIT. Collected from the SSDS, this starts at 2000 and ends at 2021. 

\begin{figure}
    \centering
    \includegraphics[width=0.90\linewidth]{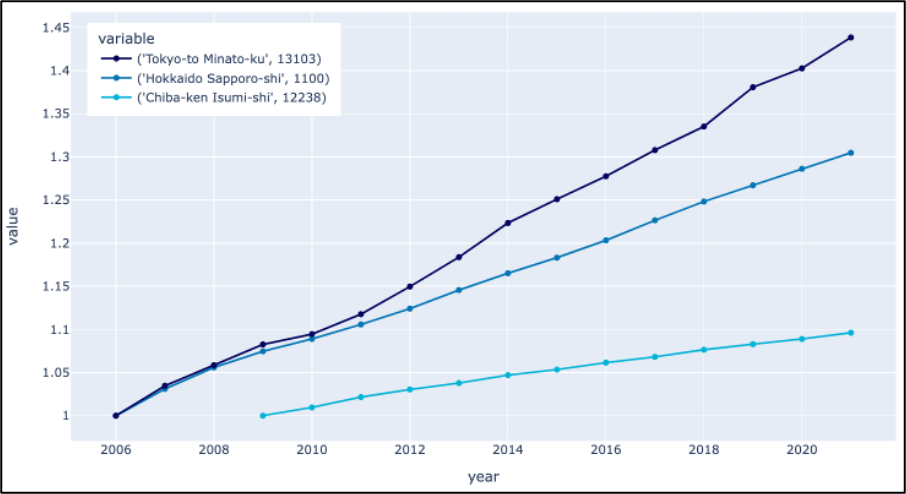}
    \caption{Cumulative new dwellings for selected municipalities}
    \label{fig:cum_new_dwellings}
\end{figure}

Once both datasets are cleaned, we interpolate the number of dwellings in the five year delay. We divide the number of new dwellings by the approximate number of existing dwellings to get the new dwellings ratio. Figure~\ref{fig:cum_new_dwellings} shows us how the number of dwellings has increased in our selected municipalities.

\subsubsection{Simple Linear Model}

\begin{figure} 
  \begin{subfigure}[b]{0.5\linewidth}
    \centering
    \frame{\includegraphics[width=0.95\linewidth]{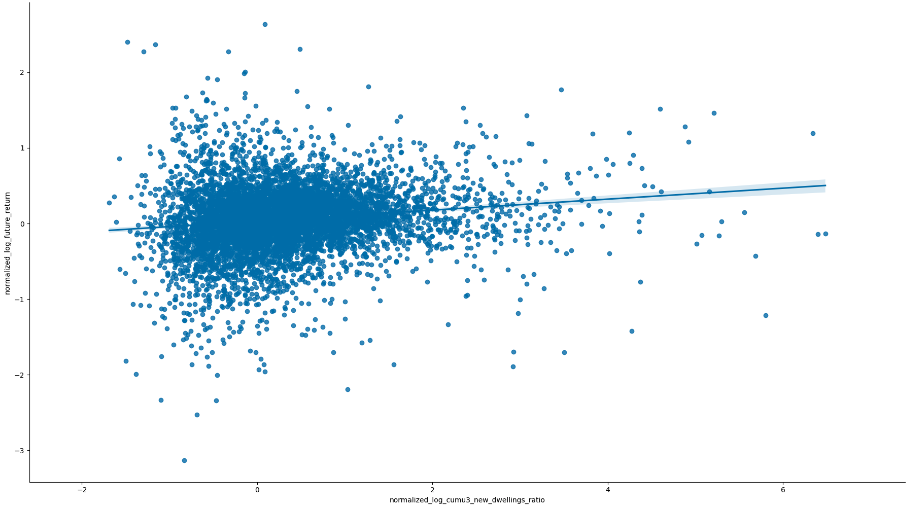}}
    \label{fig:new_dwellings_ols:a} 
    \vspace{2ex}
  \end{subfigure}
  \begin{subfigure}[b]{0.5\linewidth}
    \centering
    \frame{\includegraphics[width=0.95\linewidth]{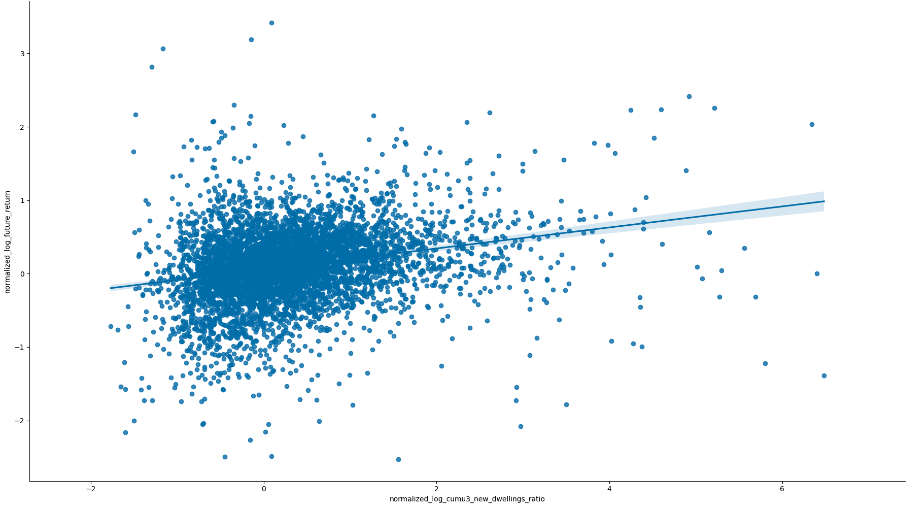}} 
    \label{fig:new_dwellings_ols:b} 
    \vspace{2ex}
  \end{subfigure} 
  \begin{subfigure}[b]{0.5\linewidth}
    \centering
    \frame{\includegraphics[width=0.95\linewidth]{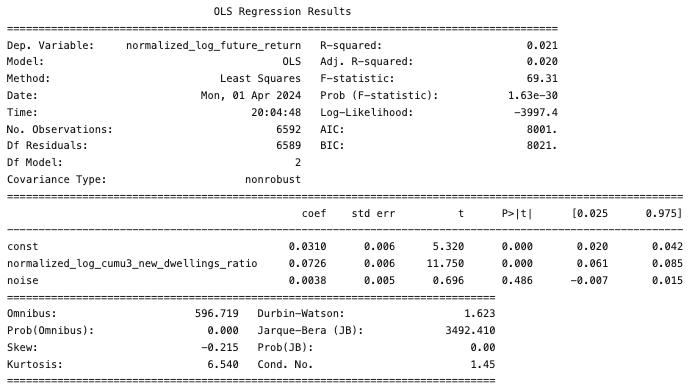}} 
    \caption{\footnotesize \textbf{Horizon: 2y} – $R^2$: 0.021; Coef: 0.0726;} 
    \label{fig:new_dwellings_ols:c} 
  \end{subfigure}
  \begin{subfigure}[b]{0.5\linewidth}
    \centering
    \frame{\includegraphics[width=0.95\linewidth]{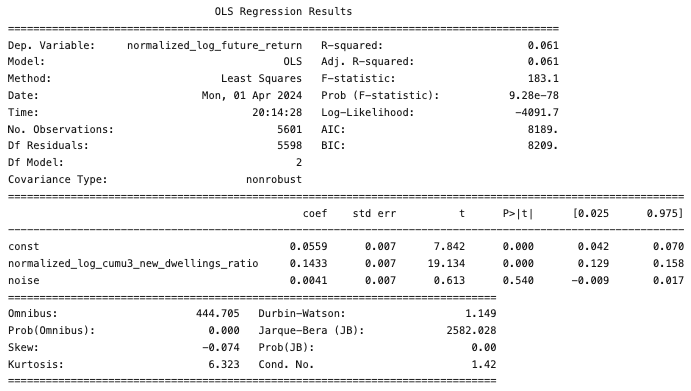}} 
    \caption{\footnotesize \textbf{Horizon: 4y} – $R^2$: 0.061; Coef: 0.1433;} 
    \label{fig:new_dwellings_ols:d} 
  \end{subfigure} 
  \caption{Results of linear model for new dwellings ratio}
  \label{fig:new_dwellings_ols} 
\end{figure}

The results of our simple regression is slightly unexpected (Figure~\ref{fig:new_dwellings_ols}). We see a positive correlation between the number of newly constructed dwellings and housing prices. Recall that our hypothesis regarding this variable required a constant demand. It is possible that developers choose markets that they predict (with a decent degree of accuracy) to have rising prices and choose to construct there. The subsequent increase in supply may work to curb the growth in prices, but is not sufficient to eliminate them. 

One might also suggest that the influx of newly constructed dwellings shift the composition of the houses sold in a period to newer, more expensive houses. While that may be true, remember that our quality adjusted price index includes age as a factor. Therefore, if the regression model is accurate, the differing composition should not affect the index.

\subsubsection{Long Short Strategy}

\begin{figure} 
  \begin{subfigure}[b]{0.5\linewidth}
    \centering
    \frame{\includegraphics[width=0.95\linewidth]{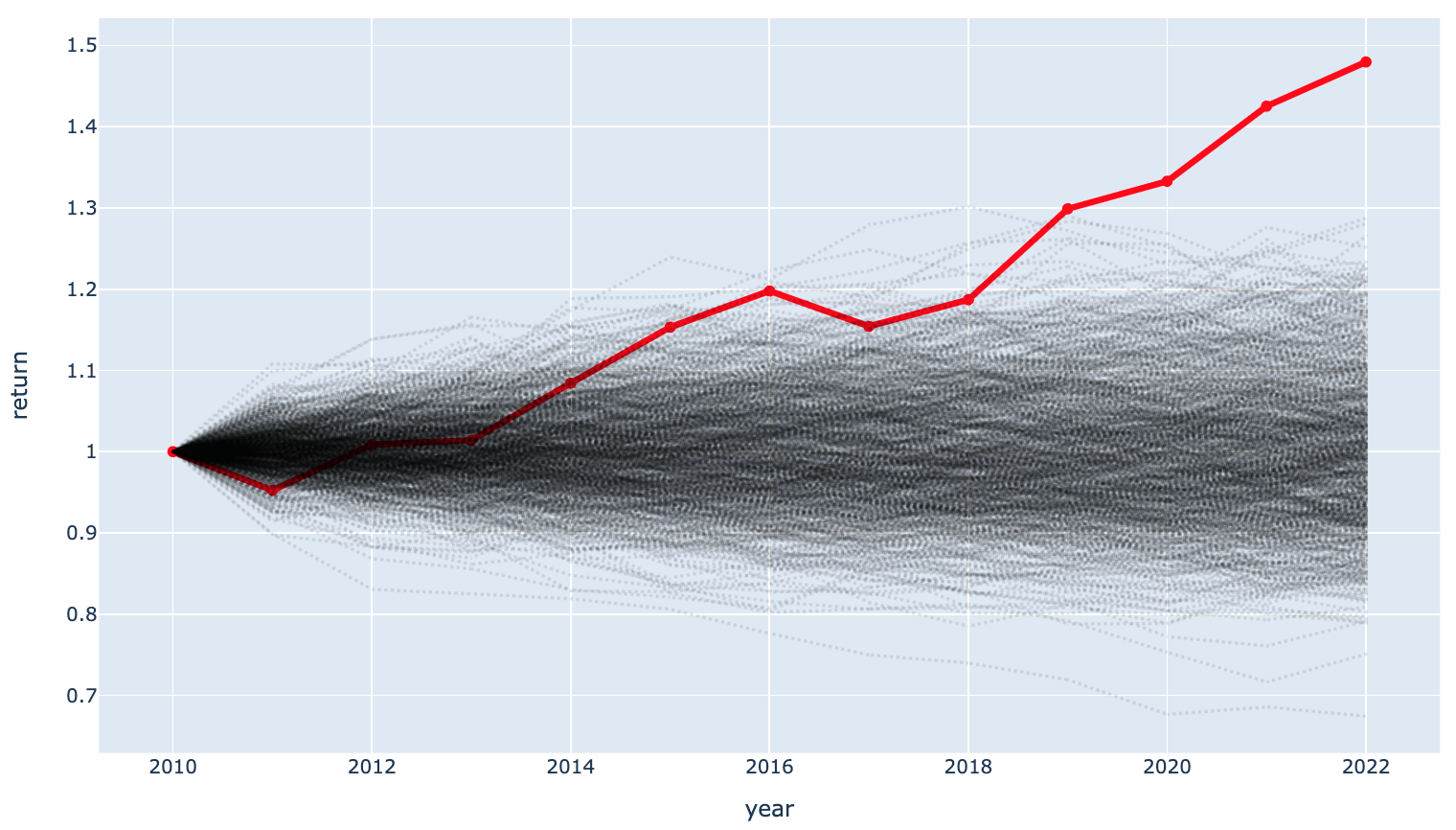}}
    \label{fig:new_dwellings_long_short:a} 
    \caption{\footnotesize \textbf{Horizon: 1y} - CAGR: 3.06\%; Sharpe: 0.83;} 
    \vspace{2ex}
  \end{subfigure}
  \begin{subfigure}[b]{0.5\linewidth}
    \centering
    \frame{\includegraphics[width=0.95\linewidth]{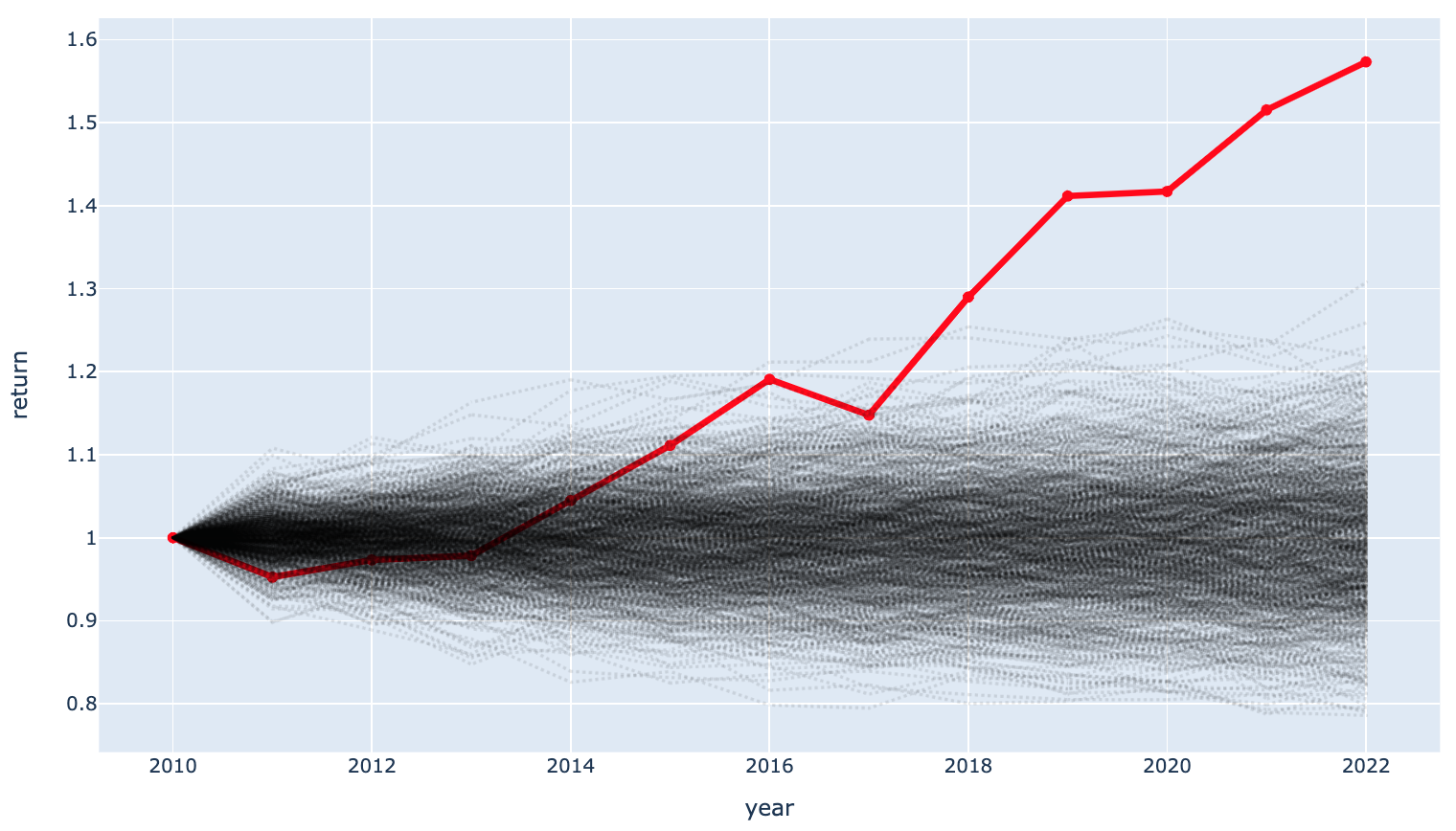}} 
    \label{fig:new_dwellings_long_short:b} 
    \caption{\footnotesize \textbf{Horizon: 2y} - CAGR: 3.54\%; Sharpe: 0.80;}
    \vspace{2ex}
  \end{subfigure} 
  \begin{subfigure}[b]{0.5\linewidth}
    \centering
    \frame{\includegraphics[width=0.95\linewidth]{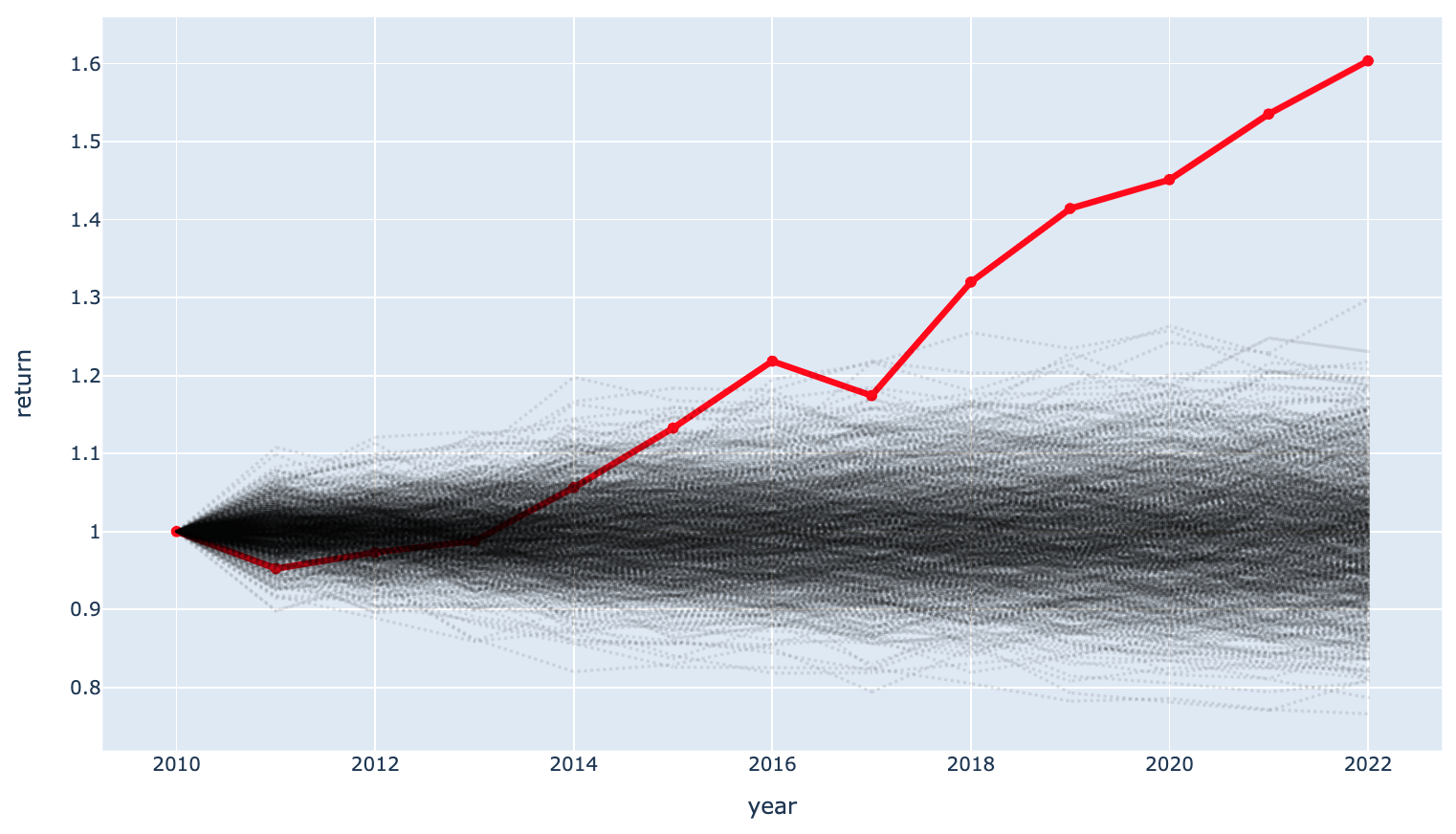}} 
    \caption{\footnotesize \textbf{Horizon: 3y} - CAGR: 3.70\%; Sharpe: 0.88;} 
    \label{fig:new_dwellings_long_short:c} 
  \end{subfigure}
  \begin{subfigure}[b]{0.5\linewidth}
    \centering
    \frame{\includegraphics[width=0.95\linewidth]{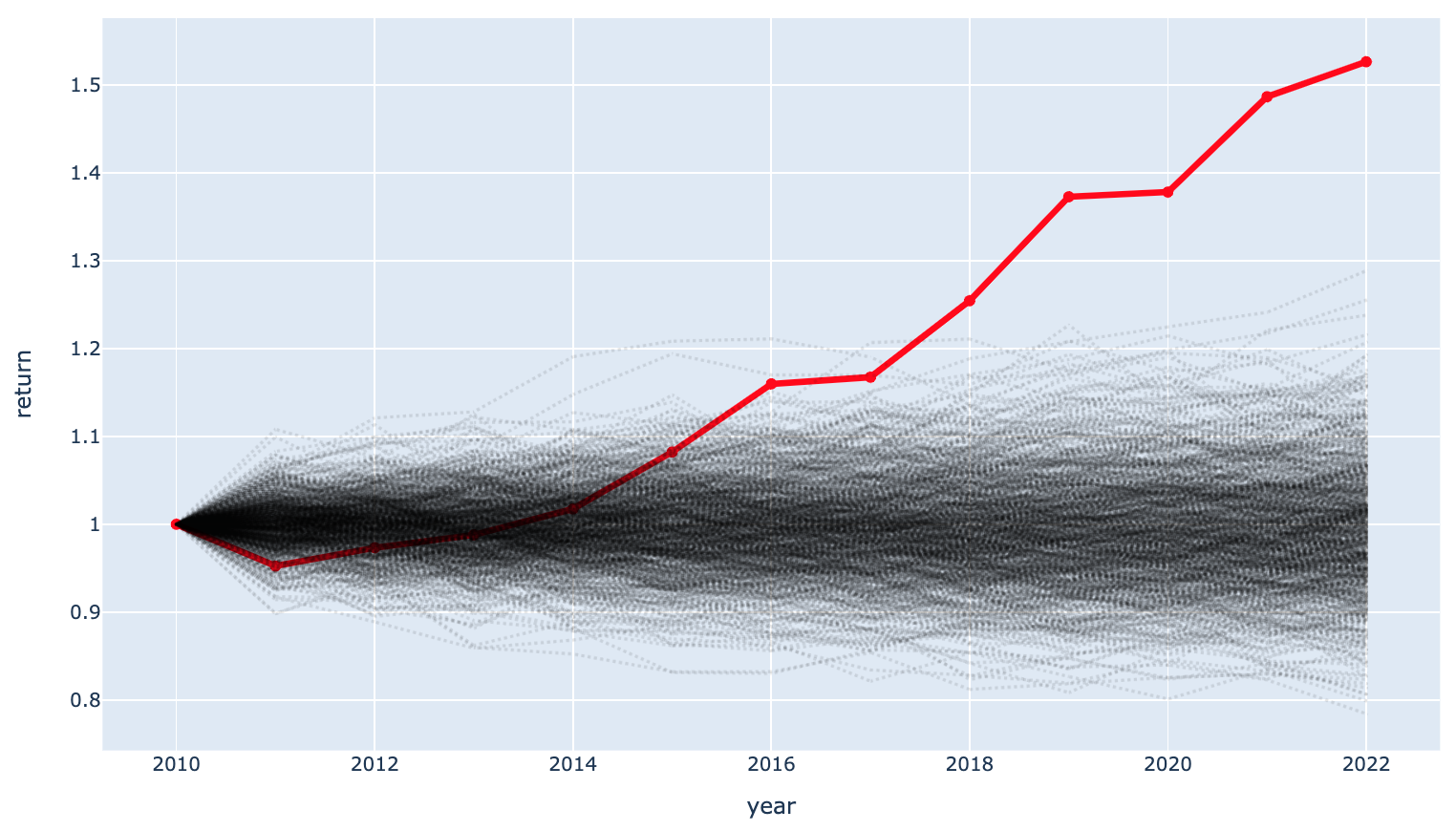}} 
    \caption{\footnotesize \textbf{Horizon: 4y} - CAGR: 3.30\%; Sharpe: 0.94;} 
    \label{fig:new_dwellings_long_short:d} 
  \end{subfigure}

  \caption{Returns from new dwellings ratio as a signal}
  \label{fig:new_dwellings_long_short} 
\end{figure}

In our strategy results (Figure~\ref{fig:new_dwellings_long_short}), we see that an increase in the number of new dwellings shows strong returns in all horizons. However, unlike net migrations and taxable income, the returns diminish between the 3rd and 4th year.

\newpage
\subsection{Historical Returns}

The concept of historical returns influencing future returns often revolves around two concepts: momentum and mean reversion. Momentum posits that periods of high returns are likely to be succeeded by similarly high returns. In contrast, mean reversion suggests that such periods will be followed by lower returns as prices adjust back toward a long-term average.

Case and Shiller’s seminal 1989 \cite{Case1988} study, as well as research by Capozza et. al. (2004) \cite{Capozza2004}, employed extensive datasets to demonstrate that the house prices exhibit significant serial correlation and mean reversion.

\subsubsection{Simple Linear Model}

\begin{figure} 
  \begin{subfigure}[b]{0.5\linewidth}
    \centering
    \frame{\includegraphics[width=0.95\linewidth]{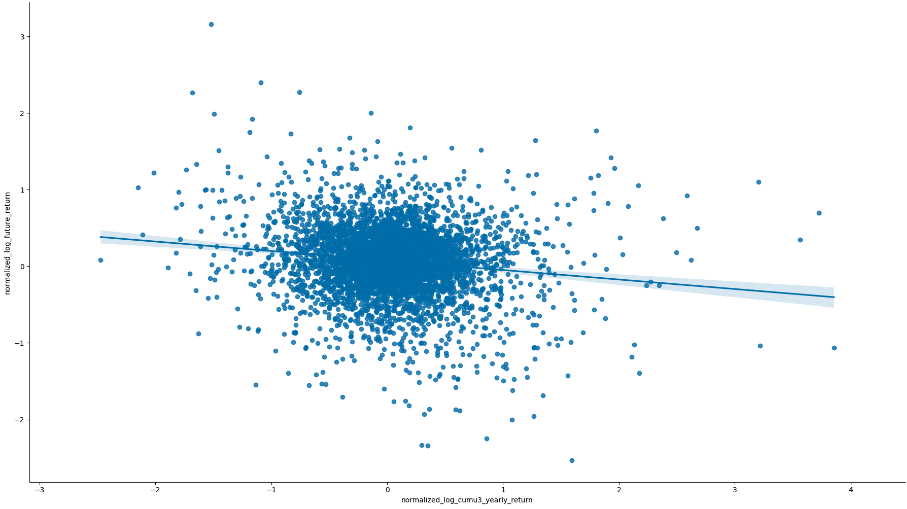}}
    \label{fig:historical_returns_ols:a} 
    \vspace{2ex}
  \end{subfigure}
  \begin{subfigure}[b]{0.5\linewidth}
    \centering
    \frame{\includegraphics[width=0.95\linewidth]{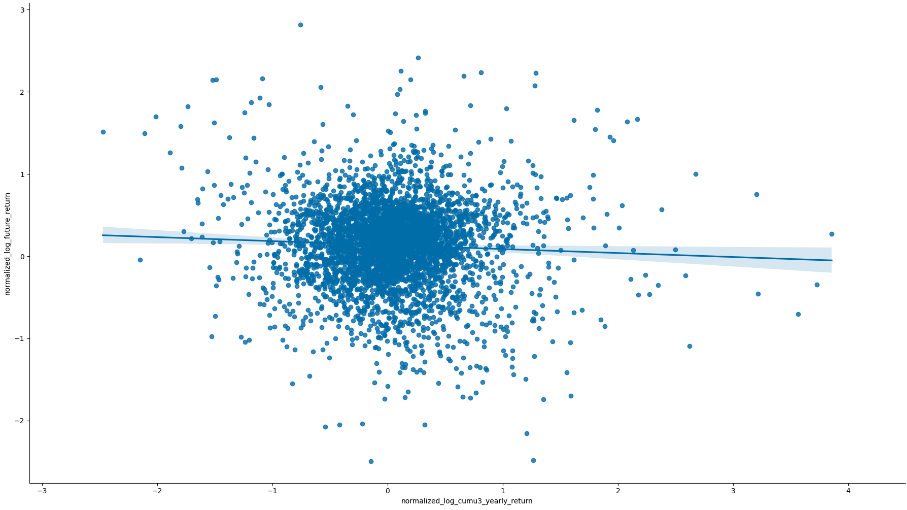}} 
    \label{fig:historical_returns_ols:b} 
    \vspace{2ex}
  \end{subfigure} 
  \begin{subfigure}[b]{0.5\linewidth}
    \centering
    \frame{\includegraphics[width=0.95\linewidth]{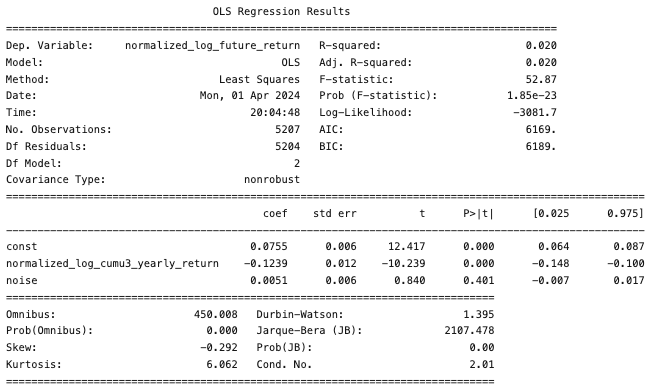}} 
    \caption{\footnotesize \textbf{Horizon: 2y} – $R^2$: 0.020; Coef: -0.1239;} 
    \label{fig:historical_returns_ols:c} 
  \end{subfigure}
  \begin{subfigure}[b]{0.5\linewidth}
    \centering
    \frame{\includegraphics[width=0.95\linewidth]{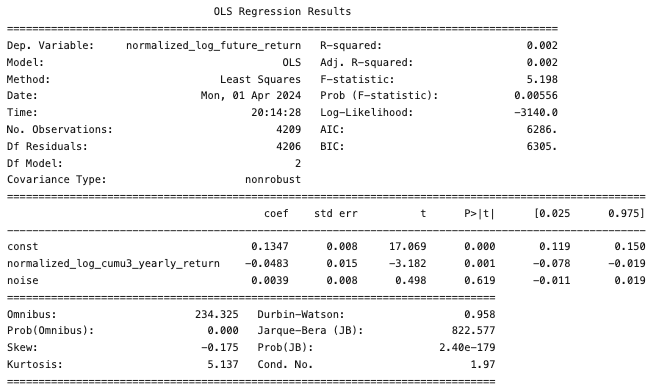}} 
    \caption{\footnotesize \textbf{Horizon: 4y} – $R^2$: 0.002; Coef: -0.0483;} 
    \label{fig:historical_returns_ols:d} 
  \end{subfigure} 
  \caption{Results of linear model for historical returns}
  \label{fig:historical_returns_ols} 
\end{figure}

For this dataset, see that the model shows a weak negative relationship between the historical cumulative return and the future returns, which decreases over a longer time horizon (Figure~\ref{fig:historical_returns_ols}). This suggests short term mean reversion.

\subsubsection{Long Short Strategy}

\begin{figure} 
  \begin{subfigure}[b]{0.5\linewidth}
    \centering
    \frame{\includegraphics[width=0.95\linewidth]{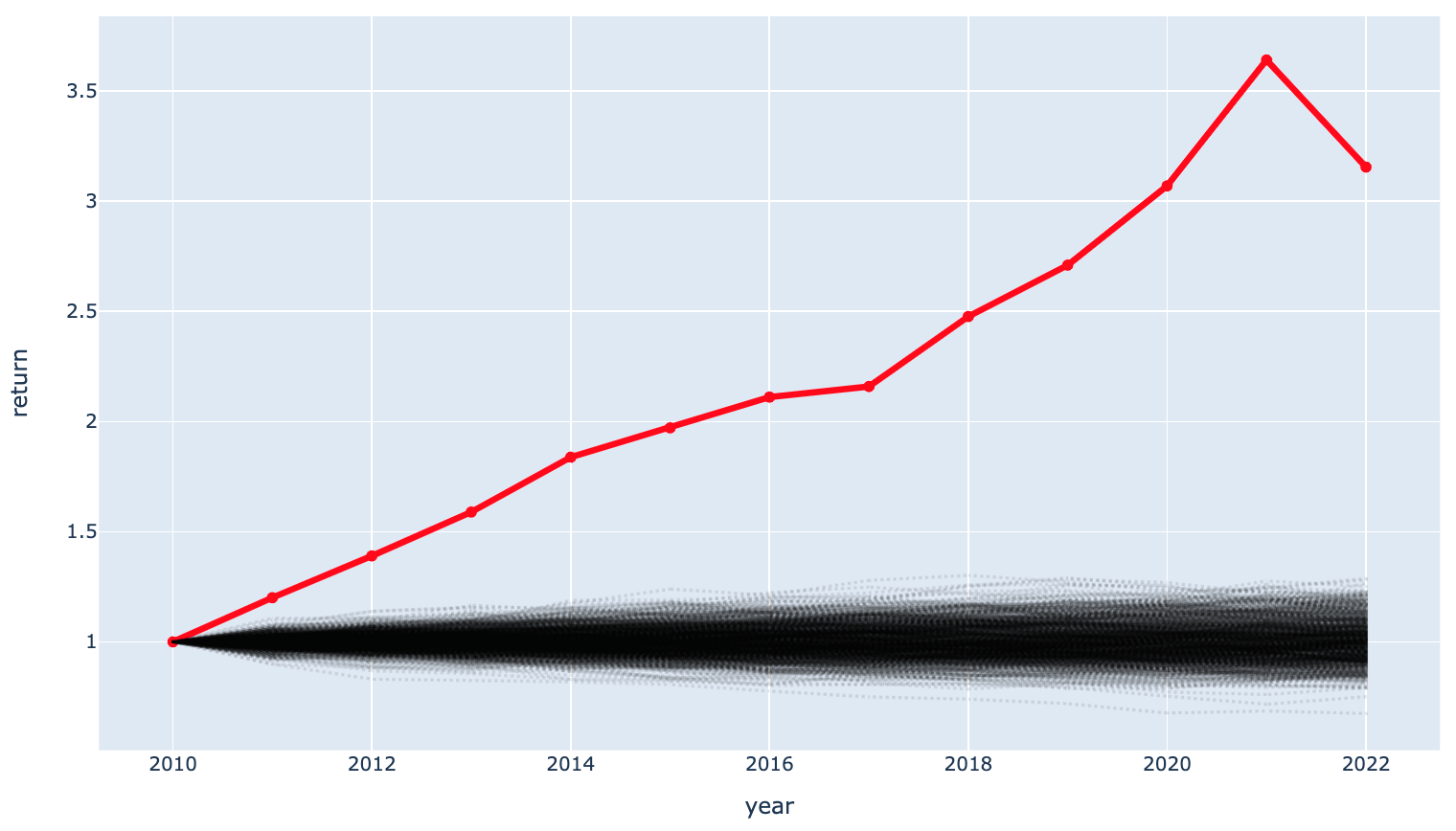}}
    \label{fig:historical_returns_long_short:a} 
    \caption{\footnotesize \textbf{Horizon: 1y} - CAGR: 9.24\%; Sharpe: 1.20;} 
    \vspace{2ex}
  \end{subfigure}
  \begin{subfigure}[b]{0.5\linewidth}
    \centering
    \frame{\includegraphics[width=0.95\linewidth]{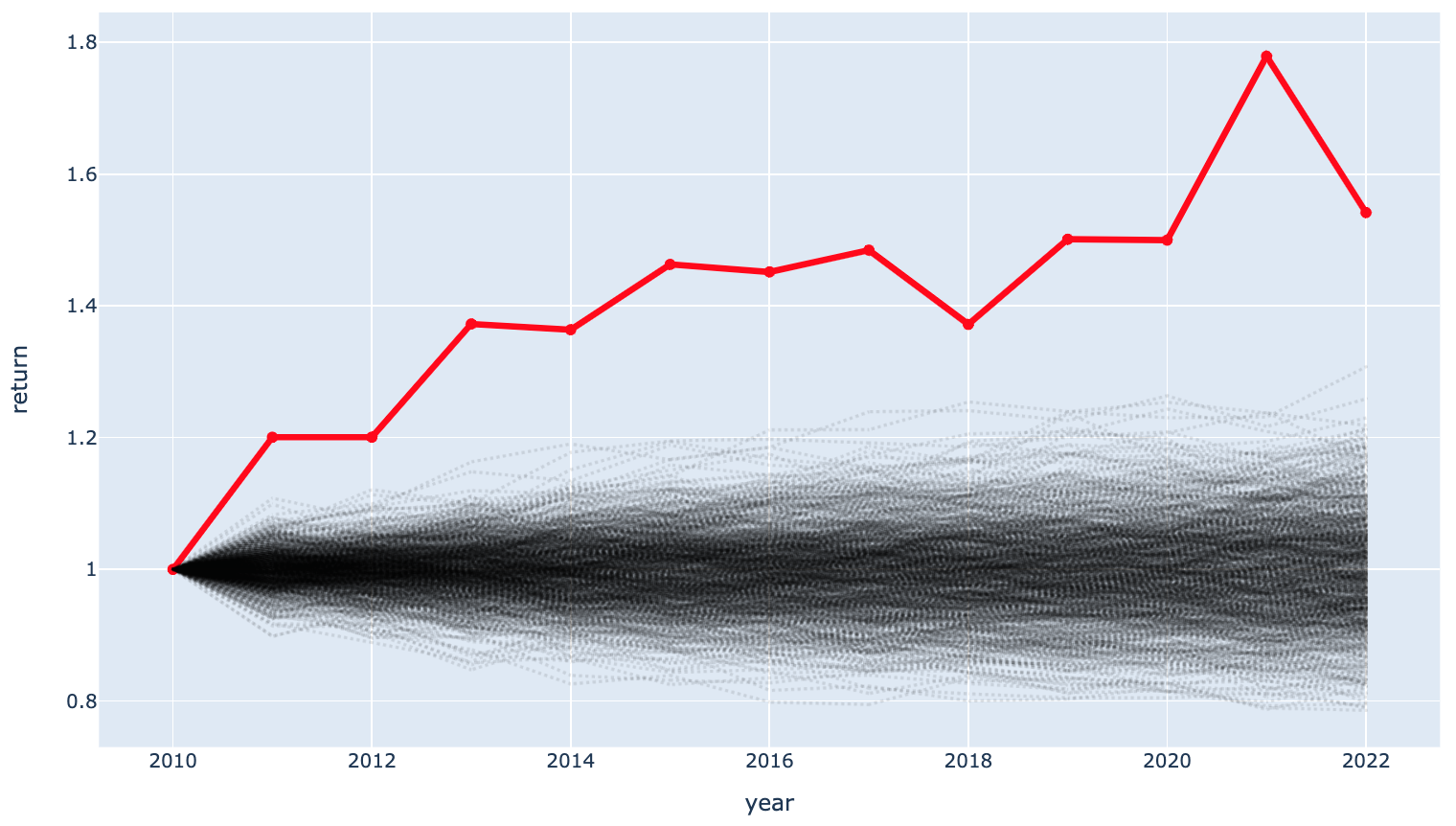}} 
    \label{fig:historical_returns_long_short:b} 
    \caption{\footnotesize \textbf{Horizon: 2y} - CAGR: 3.39\%; Sharpe: 0.42;}
    \vspace{2ex}
  \end{subfigure} 
  \begin{subfigure}[b]{0.5\linewidth}
    \centering
    \frame{\includegraphics[width=0.95\linewidth]{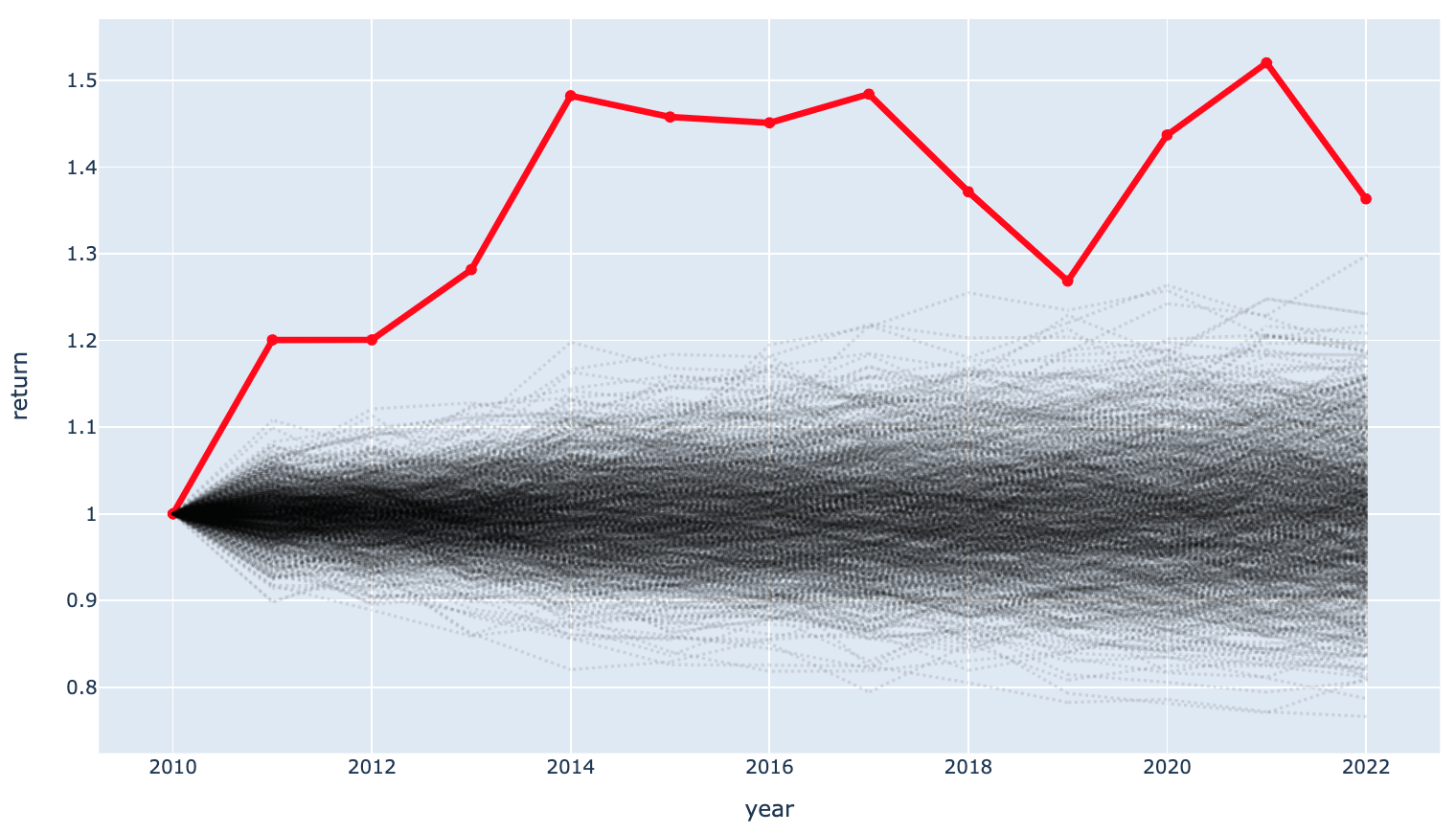}} 
    \caption{\footnotesize \textbf{Horizon: 3y} - CAGR: 2.41\%; Sharpe: 0.33;} 
    \label{fig:historical_returns_long_short:c} 
  \end{subfigure}
  \begin{subfigure}[b]{0.5\linewidth}
    \centering
    \frame{\includegraphics[width=0.95\linewidth]{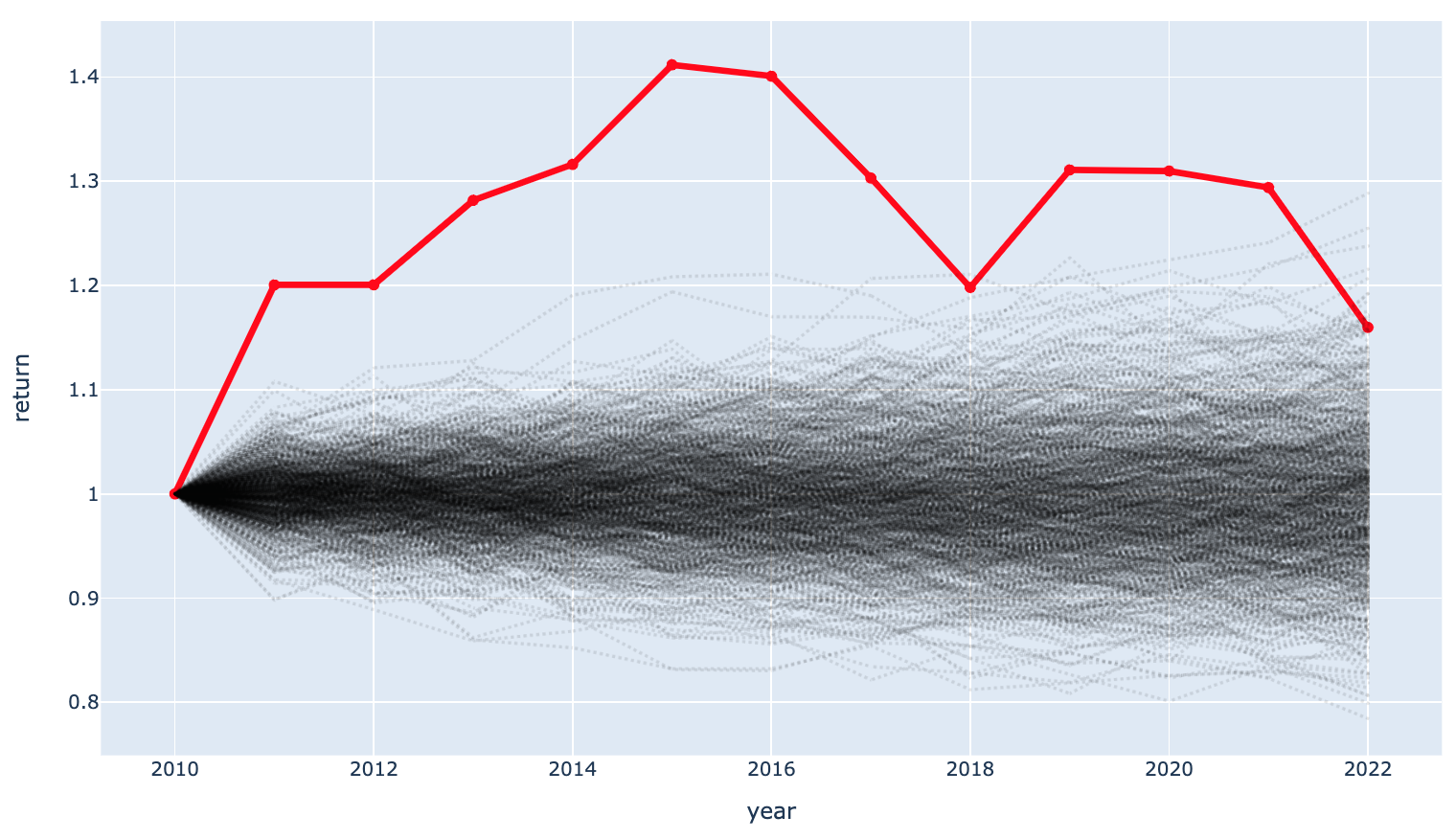}} 
    \caption{\footnotesize \textbf{Horizon: 4y} - CAGR: 1.15\%; Sharpe: 0.19;} 
    \label{fig:historical_returns_long_short:d} 
  \end{subfigure}

  \caption{Returns from inverted historical returns as a signal}
  \label{fig:historical_returns_long_short} 
\end{figure}

To try and take advantage of this apparent negative correlation, we use the inverted cumulative price growth over a trailing 3 year period as our signal. We see (Figure~\ref{fig:historical_returns_long_short}) that the mean reversion strategy demonstrates strong performance over the shortest time horizon, but its effectiveness wanes as the investment period lengthens. These simulations highlight two noteworthy observations: 

Firstly, over a time horizon of an year, the strategy does incredibly well. While the markets may exhibit mean reversion in short time spans, I suspect that a significant portion of these returns are either due to inaccuracies of the price index or seasonality. As these inaccuracies and seasonality even out over the long term, mean reversion becomes less powerful.

Second, the mean reversion strategy consistently outperforms from 2010 to 2014, regardless of the investment horizon considered. This phenomenon is likely a consequence of the market recovery following the 2008 financial crisis.

\newpage
\subsection{Spatial Price Movements}

Researchers that have tried to analyse spatial relationships for house prices consider the effect of a diffusion of information from central, important areas to nearby neighbourhoods. For example, Holly et. al. (2011) \cite{Holly2010} shows that that house prices in New York have a direct effect on the house prices in London. Such effects may exist between neighbourhoods that are in close geographical proximity in Japan. In this section, we test a simpler version of this claim. Does the weighted average return of nearby municipalities impact the future returns of an area?

\subsubsection{Dataset}

\begin{figure}
    \centering
    \includegraphics[width=0.90\linewidth]{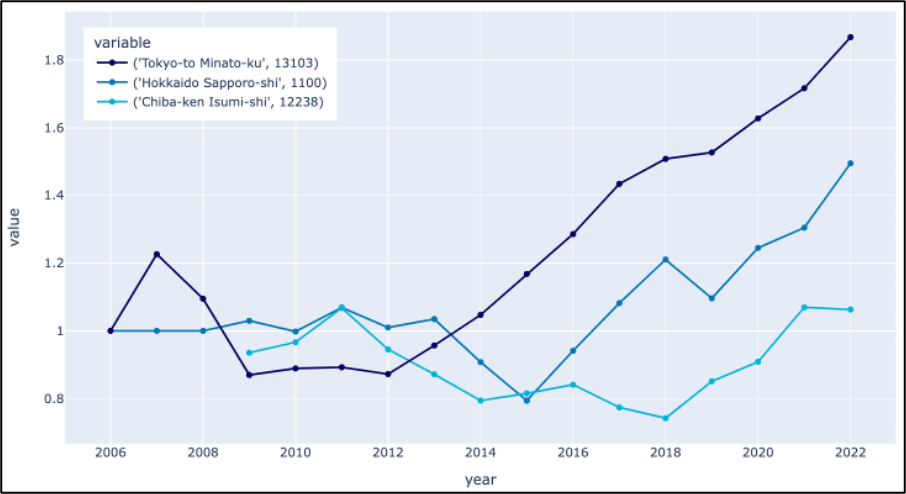}
    \caption{Cumulative neighbourhood price growth for selected municipalities}
    \label{fig:cum_neighbor_returns}
\end{figure}

To compose the dataset, we grab the 5 nearest neighbours of our municipality and take the distance weighted mean of the returns in of the neighbours in that year. Figure~\ref{fig:cum_neighbor_returns} shows these values for the three example municipalities over time.

\subsubsection{Simple Linear Model}

\begin{figure} 
  \begin{subfigure}[b]{0.5\linewidth}
    \centering
    \frame{\includegraphics[width=0.95\linewidth]{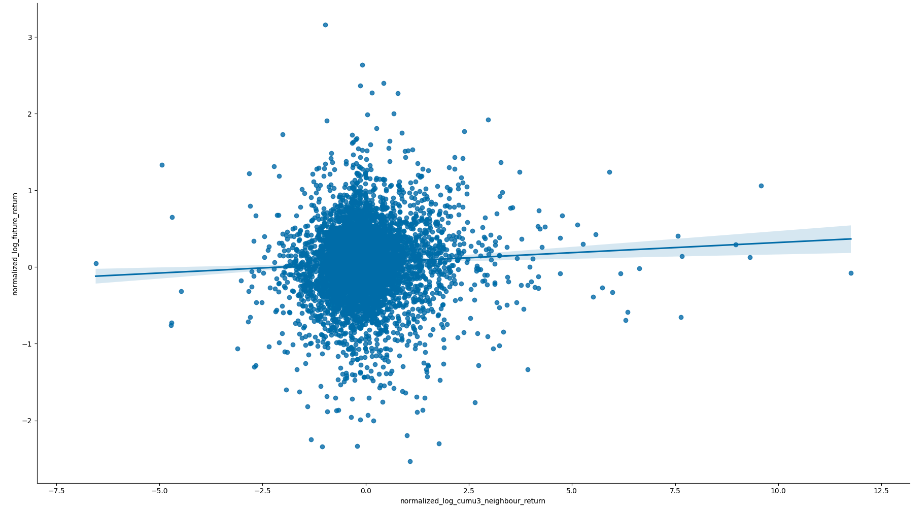}}
    \label{fig:neighbor_returns_ols:a} 
    \vspace{2ex}
  \end{subfigure}
  \begin{subfigure}[b]{0.5\linewidth}
    \centering
    \frame{\includegraphics[width=0.95\linewidth]{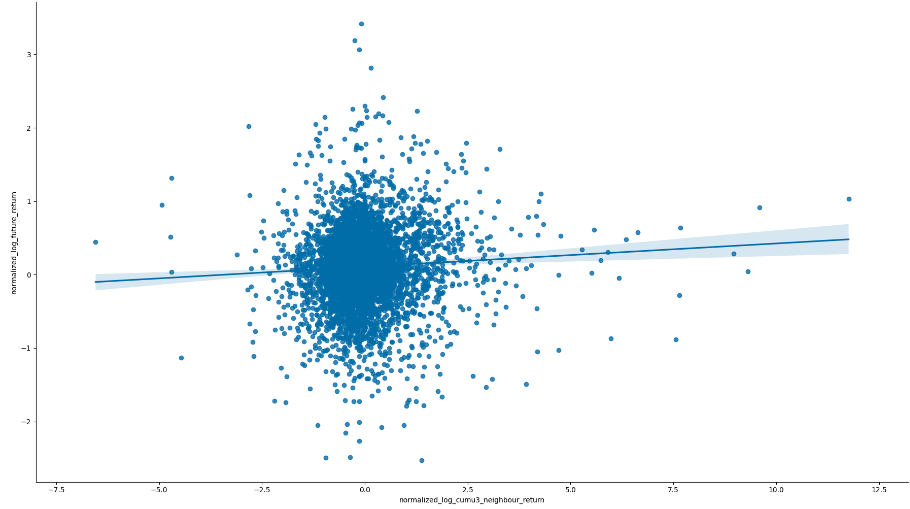}} 
    \label{fig:neighbor_returns_ols:b} 
    \vspace{2ex}
  \end{subfigure} 
  \begin{subfigure}[b]{0.5\linewidth}
    \centering
    \frame{\includegraphics[width=0.95\linewidth]{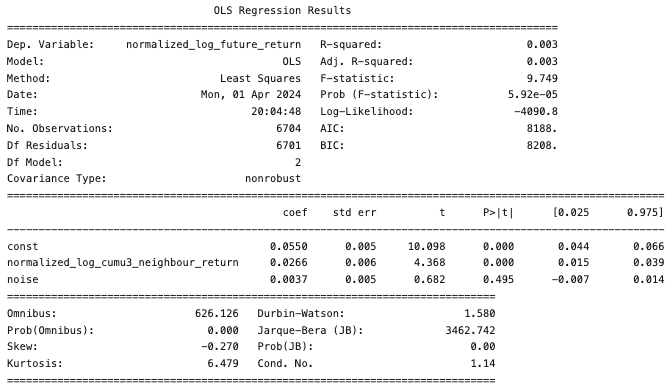}} 
    \caption{\footnotesize \textbf{Horizon: 2y} – $R^2$: 0.003; Coef: 0.0266;} 
    \label{fig:neighbor_returns_ols:c} 
  \end{subfigure}
  \begin{subfigure}[b]{0.5\linewidth}
    \centering
    \frame{\includegraphics[width=0.95\linewidth]{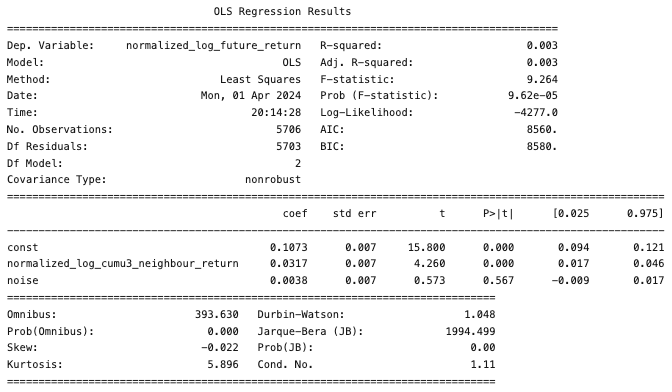}} 
    \caption{\footnotesize \textbf{Horizon: 4y} – $R^2$: 0.003; Coef: 0.0317;} 
    \label{fig:neighbor_returns_ols:d} 
  \end{subfigure} 
  \caption{Results of linear model for weighted average returns of neighbours}
  \label{fig:neighbor_returns_ols} 
\end{figure}

Our linear model for cumulative average neighbour returns does not prove capable at predicting the housing returns in a target municipality. This is evident from the low $R^2$ scores that resulted (Figure~\ref{fig:neighbor_returns_ols}).

\subsubsection{Long Short Strategy}

\begin{figure} 
  \begin{subfigure}[b]{0.5\linewidth}
    \centering
    \frame{\includegraphics[width=0.95\linewidth]{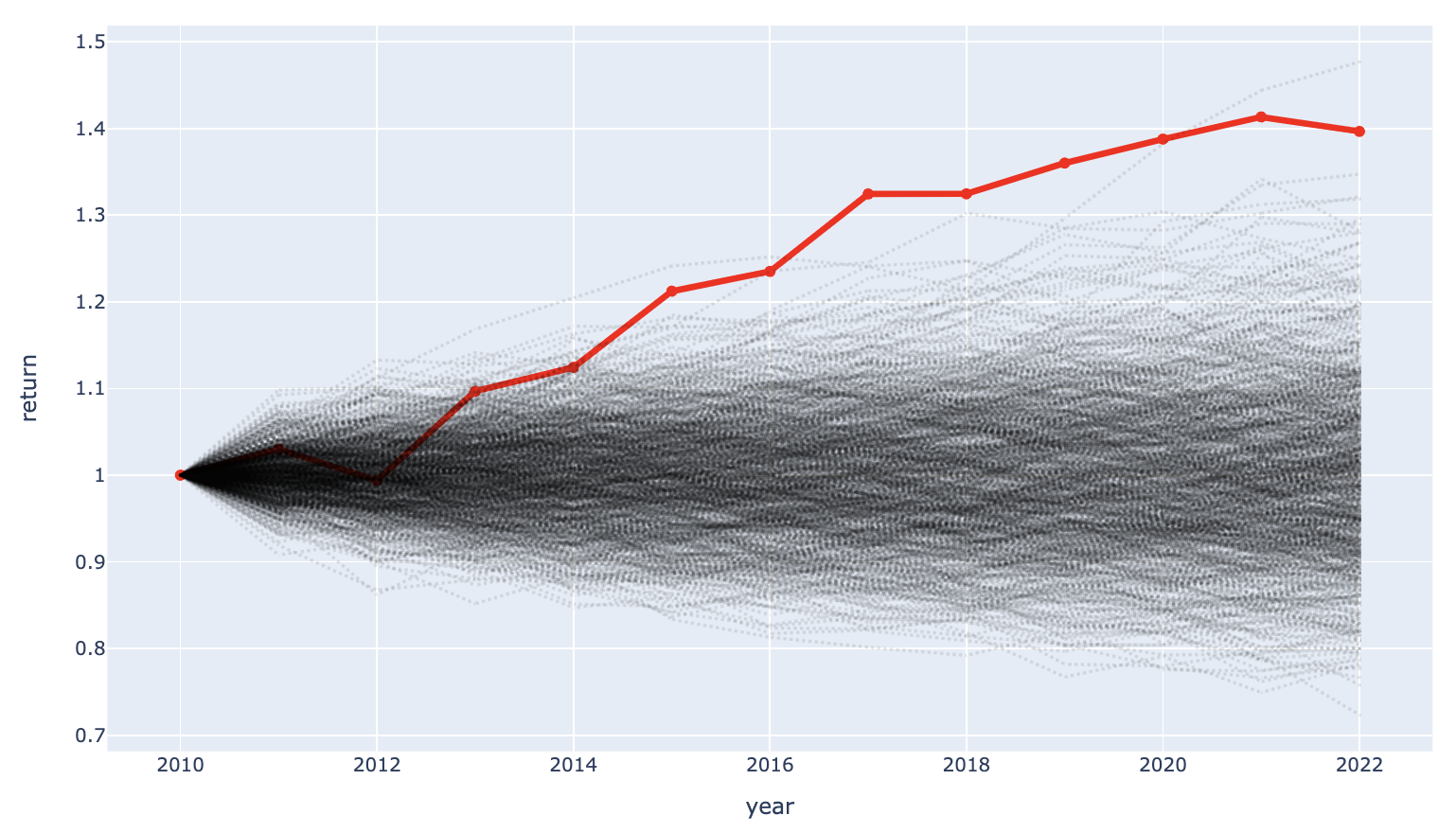}}
    \label{fig:neighbor_returns_long_short:a} 
    \caption{\footnotesize \textbf{Horizon: 1y} - CAGR: 2.60\%; Sharpe: 0.66;} 
    \vspace{2ex}
  \end{subfigure}
  \begin{subfigure}[b]{0.5\linewidth}
    \centering
    \frame{\includegraphics[width=0.95\linewidth]{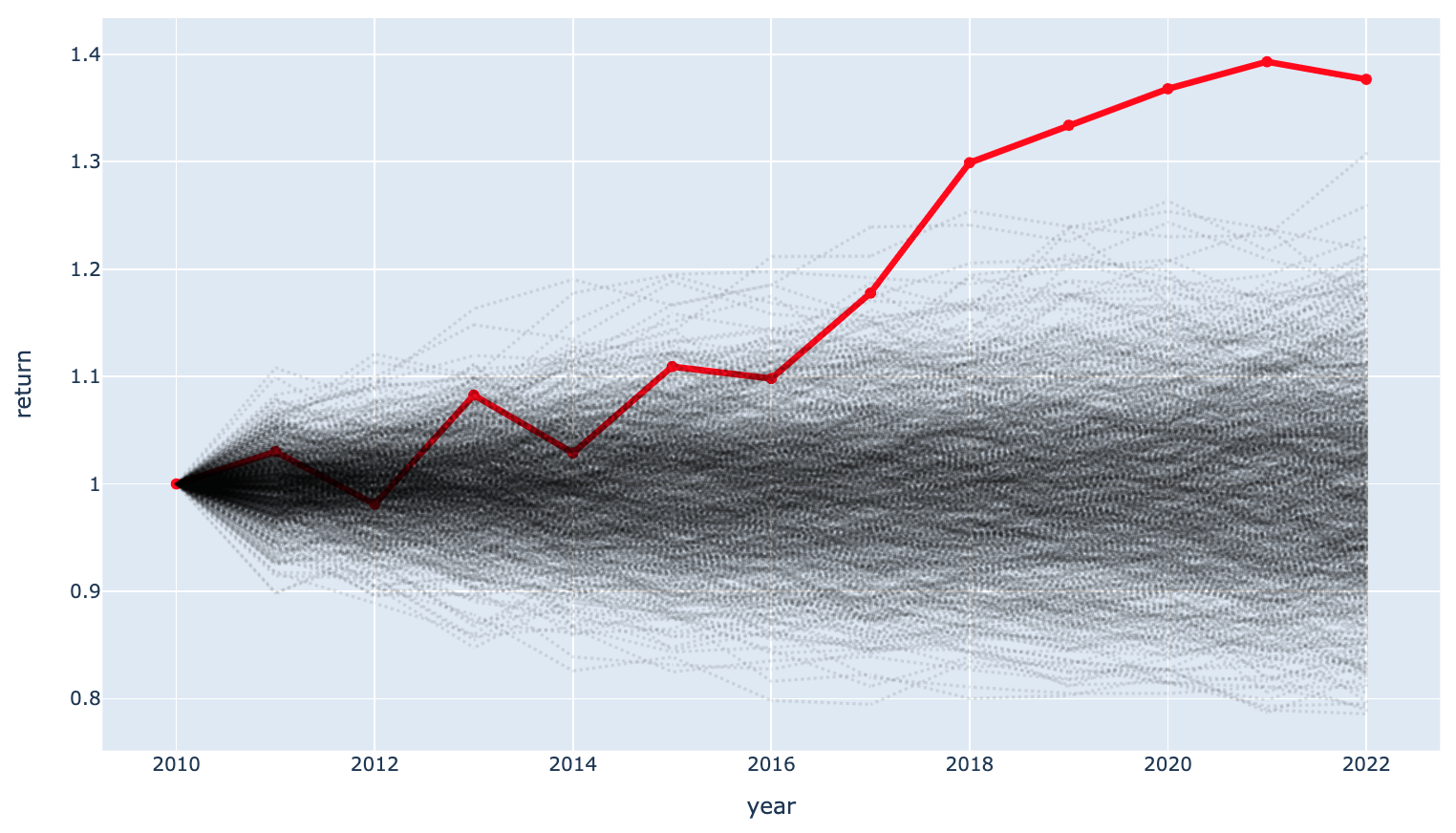}} 
    \label{fig:neighbor_returns_long_short:b} 
    \caption{\footnotesize \textbf{Horizon: 2y} - CAGR: 2.49\%; Sharpe: 0.56;}
    \vspace{2ex}
  \end{subfigure} 
  \begin{subfigure}[b]{0.5\linewidth}
    \centering
    \frame{\includegraphics[width=0.95\linewidth]{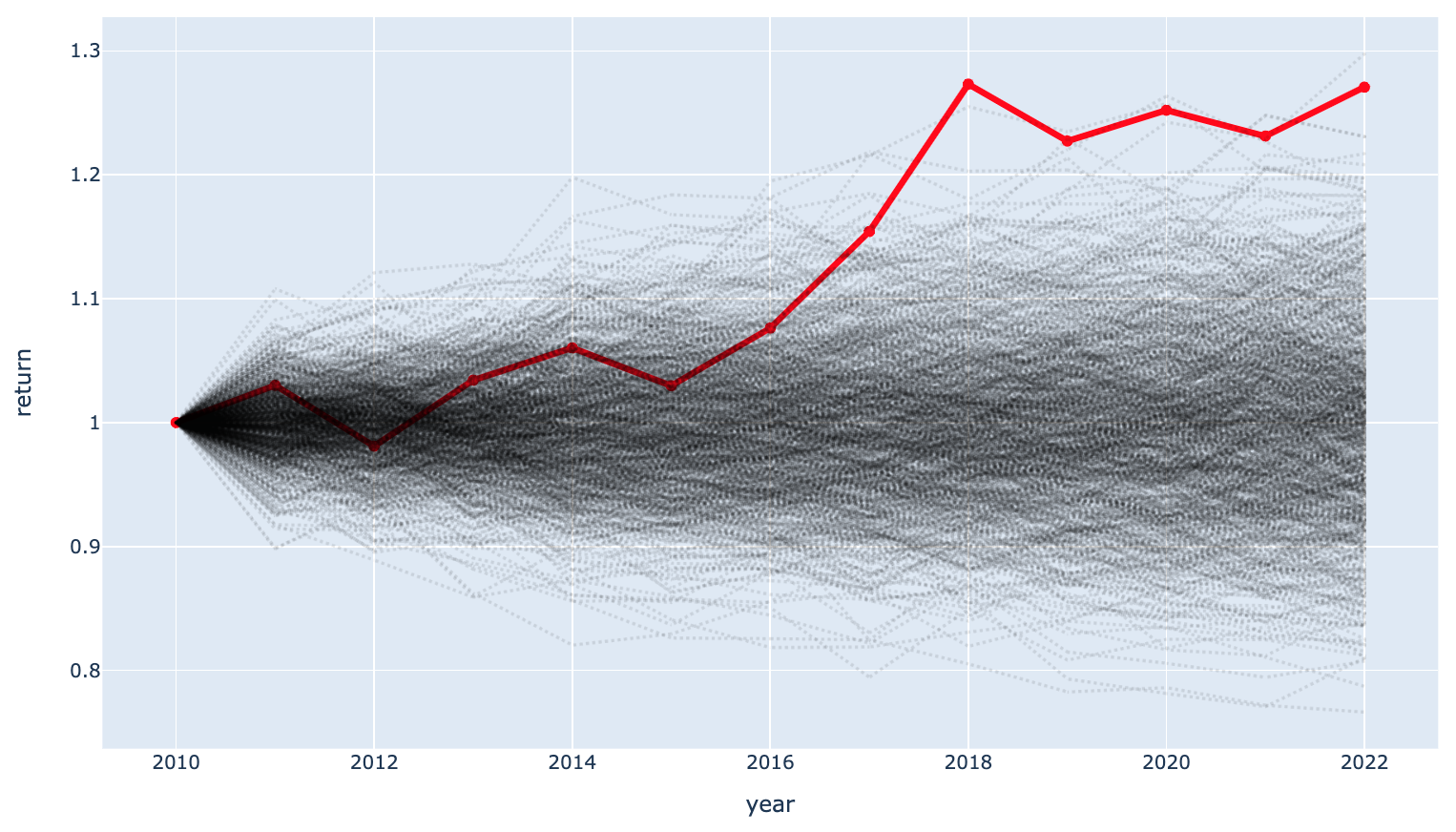}} 
    \caption{\footnotesize \textbf{Horizon: 3y} - CAGR: 1.86\%; Sharpe: 0.48;} 
    \label{fig:neighbor_returns_long_short:c} 
  \end{subfigure}
  \begin{subfigure}[b]{0.5\linewidth}
    \centering
    \frame{\includegraphics[width=0.95\linewidth]{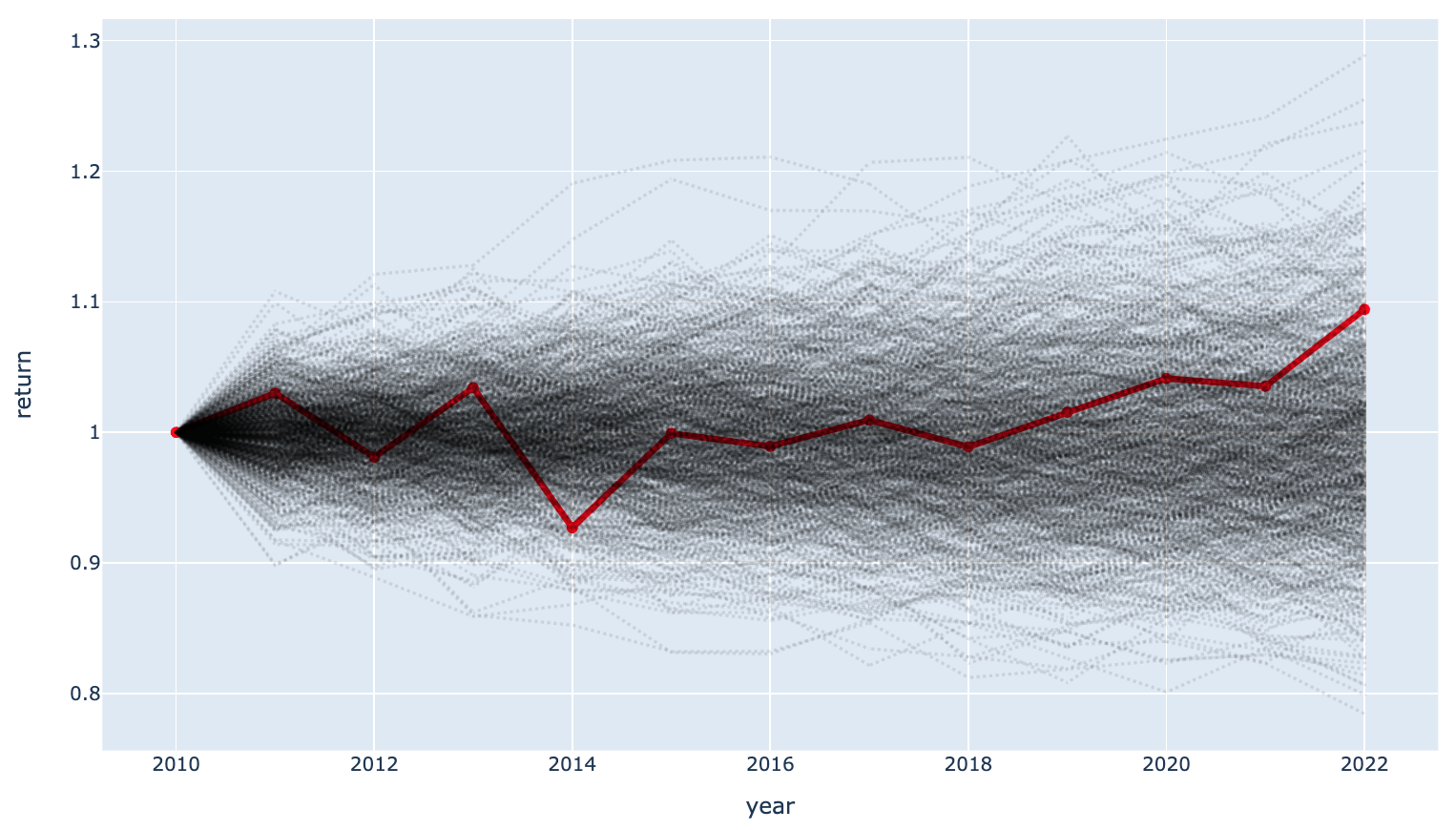}} 
    \caption{\footnotesize \textbf{Horizon: 4y} - CAGR: 0.69\%; Sharpe: 0.18;} 
    \label{fig:neighbor_returns_long_short:d} 
  \end{subfigure}

  \caption{Returns from weighted average returns of neighbours as a signal}
  \label{fig:neighbor_returns_long_short} 
\end{figure}

Of all the factors we have tested for so far, the average returns of a municipality’s closest neighbours seems to be the least predictive, and any predictive ability that is demonstrated over shorter horizons quickly diminishes to noise over longer periods (Figure~\ref{fig:neighbor_returns_long_short}).

A different, possibly more interesting study would be to use the price movements of the nearest large metropolitan as a factor, instead of a weighted average of a region’s neighbours.

\subsection{Future Work}

The search for alternative data is an ongoing endeavour crucial for transforming this project into a viable investment strategy. The datasets provided by the SSDS merely scratch the surface of what is needed. To truly unlock the predictive potential, a significant amount of work must be dedicated to identifying and acquiring datasets that can enhance the accuracy and reliability of our models. The current datasets leave much to be desired, and expanding our data arsenal would significantly increase our ability to make judgements regarding the price dynamics of a location.

\newpage
\section{Model} \label{model}

This experiment was born out of my interest in using Transformer models for time series analysis. While model architectures are undeniably impressive, it's crucial to recognize that the quality of the data at our disposal is often the limiting factor, especially for this application.

\subsection{Time Series Transformer}

Originally designed for neural machine translation by Vaswani et al. in 2017 \cite{vaswani2017attention}, Transformers have proven to be powerful, general-purpose trainable computers. A time series, like a sentence, is a sequence of numerical observations or tokens with a numerical representation. The attention mechanism that makes Transformers so effective for language modelling can also be incredibly valuable for time series forecasting.

Compared to simple linear models, Transformers offer a significant advantage in encoding temporal features. With linear models, feature engineering moving averages and cumulative values would be challenging, and each new feature would introduce a hyperparameter to tune. In contrast, Transformers can create any relevant representations of the data, making these hyperparameters trainable.

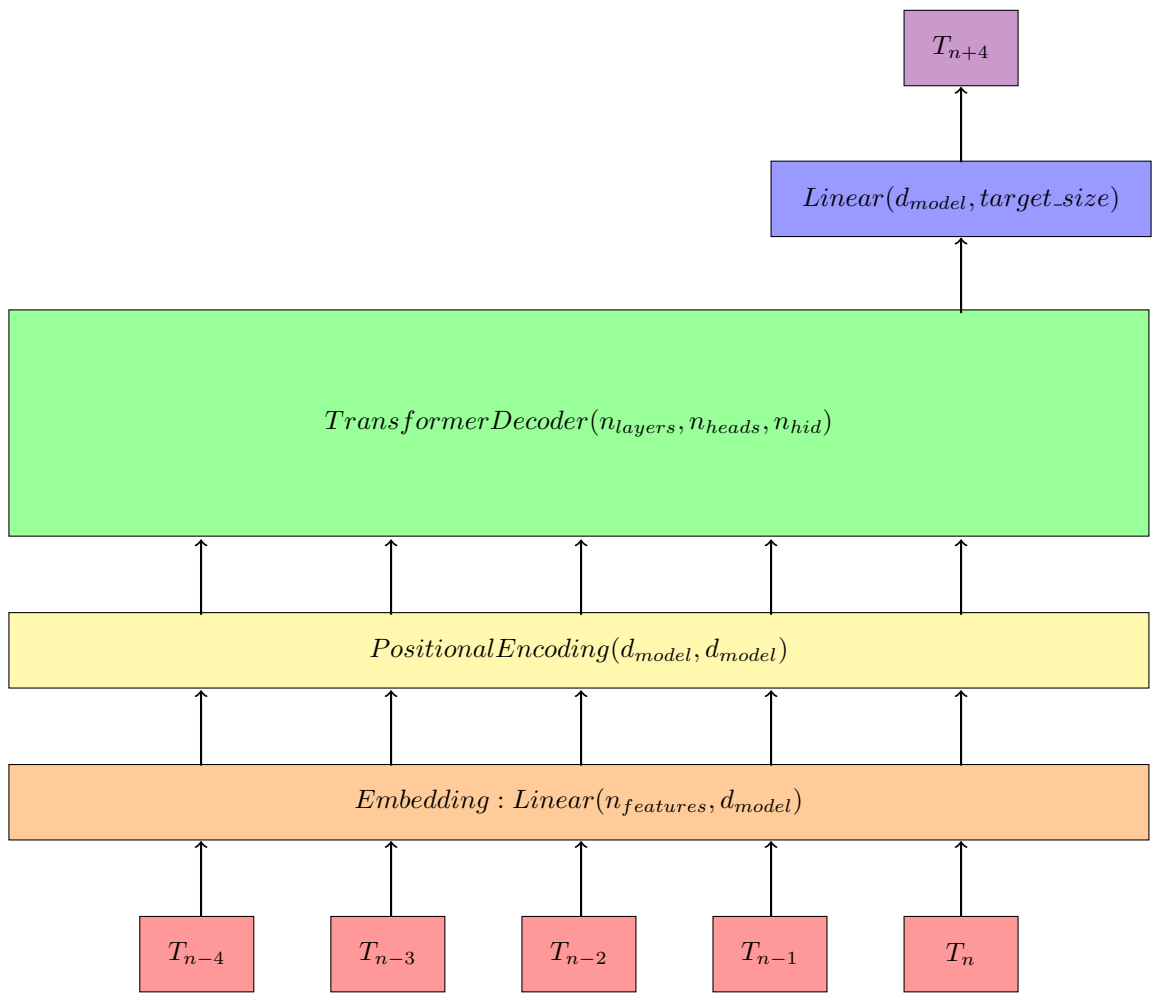
\begin{figure}
    \centering
    \begin{tikzpicture}[
        thick, text centered,
        box/.style={draw, thin, minimum width=1.5cm, minimum height=1cm},
        mid_wide_box/.style={draw, thin, minimum width=5cm, minimum height=1cm},
        wide_box/.style={draw, thin, minimum width=15cm, minimum height=1cm},
        thick_wide_box/.style={draw, thin, minimum width=15cm, minimum height=3cm},
      ]
co
      \node[box, fill=red!40] (tn) {$T_n$};
      \node[box, fill=red!40, left=1 of tn] (tnm1) {$T_{n-1}$};
      \node[box, fill=red!40, left=1 of tnm1] (tnm2) {$T_{n-2}$};
      \node[box, fill=red!40, left=1 of tnm2] (tnm3) {$T_{n-3}$};
      \node[box, fill=red!40, left=1 of tnm3] (tnm4) {$T_{n-4}$};
    
      \node[wide_box, fill=orange!40, above=1 of tnm2] (embedding) {$Embedding: Linear(n_{features}, d_{model})$} ;
      \node[wide_box, fill=yellow!40, above=1 of embedding] (positional) {$PositionalEncoding(d_{model}, d_{model})$};
      \node[thick_wide_box, fill=green!40, above=1 of positional] (decoder) {$TransformerDecoder(n_{layers}, n_{heads}, n_{hid})$};
      
      \node[mid_wide_box, fill=blue!40, above=9 of tn] (linear){$Linear(d_{model}, target\_size)$};

      \node[box, fill=violet!40, above=11 of tn] (xnp4) {$T_{n+4}$};
      
      \draw[->] (-0.0,0.5) -- (-0.0,1.5);
      \draw[->] (-2.5,0.5) -- (-2.5,1.5);
      \draw[->] (-5.0,0.5) -- (-5.0,1.5);
      \draw[->] (-7.5,0.5) -- (-7.5,1.5);
      \draw[->] (-10.0,0.5) -- (-10.0,1.5);
    
      \draw[->] (-0.0,2.5) -- (-0.0,3.5);
      \draw[->] (-2.5,2.5) -- (-2.5,3.5);
      \draw[->] (-5.0,2.5) -- (-5.0,3.5);
      \draw[->] (-7.5,2.5) -- (-7.5,3.5);
      \draw[->] (-10.0,2.5) -- (-10.0,3.5);
    
      \draw[->] (-0.0,4.5) -- (-0.0,5.5);
      \draw[->] (-2.5,4.5) -- (-2.5,5.5);
      \draw[->] (-5.0,4.5) -- (-5.0,5.5);
      \draw[->] (-7.5,4.5) -- (-7.5,5.5);
      \draw[->] (-10.0,4.5) -- (-10.0,5.5);
    
      \draw[->] (-0.0,8.5) -- (-0.0,9.5);
      
      \draw[->] (-0.0,10.5) -- (-0.0,11.5);
    
    \end{tikzpicture}
    \caption{Time series transformer}
    \label{fig:time_series_transformer}
\end{figure}

The process starts by applying positional encoding to the input sequence, adding information about the relative position of each vector. These positionally encoded embeddings are then passed through the Transformer decoder. The decoder's final output is extracted and fed into a linear layer, producing the desired result. Figure~\ref{fig:time_series_transformer} illustrates the key components of our transformer model and how temporal input data would flow through it.

\subsection{Time Series Dataset}

Our time series dataset includes the target variable and factors over a window spanning up to the last 5 years.

\subsubsection{Target}

The goal is to predict the future relative risk-adjusted return for a given location. We've calculated the risk-adjusted returns for each period over the past four years and normalized these values annually. The model will predict the relative risk-adjusted returns from any time point t to t+4

\subsubsection{Features}

We merge our datasets, all indexed by area code and year, to create a comprehensive dataset (Table~\ref{tab:combined_dataset}).

\begin{table}[ht]
\centering
\begin{tabular}{ccccccccc}
\toprule
Year & Area Code & Target & Taxable Income Growth & Net Migration Ratio & \ldots \\
\midrule
2007 & 13101 & 0.858 & 0.107 & 0.009 & \ldots \\
2008 & 13101 & 0.341 & 0.069 & 0.011 & \ldots \\
2009 & 13101 & -0.587 & -0.087 & 0.020 & \ldots \\
\ldots & \ldots & \ldots & \ldots & \ldots & \ldots \\
2019 & 13101 & -0.486 & 0.138 & 0.028 & \ldots \\
2020 & 13101 & 0.308 & -0.029 & 0.015 & \ldots \\
2021 & 13101 & 0.578 & -0.007 & -0.001 & \ldots \\
\bottomrule
\end{tabular}
\caption{Prices and factors combined into a single dataset}
\label{tab:combined_dataset}
\end{table}

\subsubsection{Temporal Features}

Each input contains a data window used to predict the output. The selected look-back period for this experiment is 5 years. The shape of our window equals 5 x  $n_{\text{features}}$. Table~\ref{tab:temporal_dataset_form} demonstrates what individual samples of our data looks like, while Table~\ref{tab:single_temporal_input} shows the input for a single sample that would be used by the model for a prediction.

\begin{table}[ht]
\centering
\begin{tabular}{ccc}
\toprule
Area Code & X (window) & Y \\
\midrule
A & '09 Data & '11 Return \\
B & '09 Data & '11 Return \\
A & '09, '10 Data & '12 Return \\
B & '09, '10 Data & '12 Return \\
A & '09, '10, '11 Data & '13 Return \\
B & '09, '10, '11 Data & '13 Return \\
A & '09, '10, '11, '12 Data & '14 Return \\
B & '09, '10, '11, '12 Data & '14 Return \\
\bottomrule
\end{tabular}
\caption{Form of temporal dataset}
\label{tab:temporal_dataset_form}
\end{table}

\begin{table}[ht]
\centering
\begin{tabular}{ccccc}
\toprule
Yearly Return & Taxable Income Growth & Net Migration Ratio & New Dwellings Ratio & \ldots \\
\midrule
0.625 & 0.055  & 0.025 & 0.028 & \ldots \\
0.388 & 0.082  & 0.032 & 0.071 & \ldots \\
0.430 & 0.082  & 0.033 & 0.024 & \ldots \\
0.422 & 0.076  & 0.024 & 0.045 & \ldots \\
-0.002 & 0.108 & 0.024 & 0.018  & \ldots \\
\bottomrule
\end{tabular}
\caption{Temporal inputs from 2013 to 2017 to predict risk adjusted return till 2021 }
\label{tab:single_temporal_input}
\end{table}

\subsubsection{Spatial Features}

As discussed in Chapter~\ref{factors}, changes in factors of nearby municipalities may affect the prices of the target municipality. To account for this, we compute the n nearest neighbours for each municipality, get a data window for them (possibly with a subset of features), and join these windows to our original window along with the Euclidean distance between the target municipality and its neighbours

The window shape is $5 \times n_{\text{spatial\_features}}$. Where,

\begin{equation}
    n_{\text{spatial\_features}} = n_{\text{features}} + n_{\text{neighbours}} \times (n_{\text{neighbour\_features}} + 1)
\end{equation}

Table~\ref{tab:single_spatio_temporal_input} shows how an features of arbitrary number of neighbouring municipalities can be horizontally appended to our input dataset.

\begin{table}[ht]
\centering

\begin{minipage}[c]{0.3\linewidth}
\centering
\small
\begin{tabular}{ccc}
\toprule
Year & Target & Factor \\
\midrule
2015 & 0.625 & 1.095 \\
2016 & 0.388 & 1.111 \\
2017 & 0.43  & 1.13  \\
2018 & 0.422 & 1.151 \\
2019 & -0.002 & 1.166 \\
\bottomrule
\end{tabular}
\caption*{Target Area}
\end{minipage}
\begin{minipage}[c]{0.05\linewidth}
\centering
$+$
\end{minipage}
\begin{minipage}[c]{0.3\linewidth}
\centering
\small
\begin{tabular}{ccc}
\toprule
Year & Distance & Factor \\
\midrule
2015 & 0.6 & 0.055 \\
2016 & 0.6 & 0.082 \\
2017 & 0.6 & 0.082 \\
2018 & 0.6 & 0.076 \\
2019 & 0.6 & 0.108 \\
\bottomrule
\end{tabular}
\caption*{Neighbour 1}
\end{minipage}
\begin{minipage}[c]{0.05\linewidth}
\centering
$+\hspace{15pt}\ldots$
\end{minipage}

\caption{Example input of spatio-temporal dataset}
\label{tab:single_spatio_temporal_input}

\end{table}

\subsubsection{Weighted Cost Function}

Since the house price index is derived from transactions in an area, the estimated price index for a populous municipality like Tokyo Minato is likely more accurate than that of a small town. We use a weighted cost function to account for this, with the weight derived from the municipality's population:

\begin{equation}
    \text{weight} = 1 + \log_{10}(\text{Population})
\end{equation}

\begin{figure}
    \centering
    \includegraphics[width=0.5\linewidth]{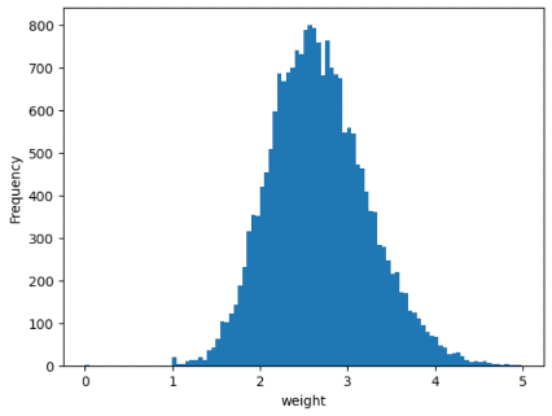}
    \caption{Distribution of weights}
    \label{fig:weight_distribution}
\end{figure}

The distribution of our weights is displayed in Figure~\ref{fig:weight_distribution}.

\subsubsection{Train Test Split}

The data was split into two segments: 2005 to 2020 for training and 2021 to 2022 for testing. A tuning set was not used since we did not extensively tune the hyperparameters. The experiment's goal was to see if the model could learn patterns to identify profitable municipalities for investment, rather than producing the most performant model.

It's important to note that the nature of financial markets makes creating a good train-test split challenging. Unlike ordinary machine learning tasks, market environments constantly fluctuate, and the post-COVID market environment likely differs significantly from preceding environments. To address this, the target variable was normalized across all regions yearly, aiming for the model to recognize relative differences in factors and risk-adjusted returns for each municipality, rather than the general macro environment impacting all municipalities.

\subsection{Results}

\subsubsection{Model Hyperparameters}
The model was configured with the following hyperparameters:
\begin{itemize}
    \item Learning Rate: $3 \times 10^{-4}$
    \item Weight Decay: 1
    \item Dropout: 0.1
    \item Number of decoder layers: 4
    \item Number of attention heads per layer: 4
    \item Dimension of hidden units: 128
    \item Dimension of embedding: 128
\end{itemize}

\subsubsection{Performance Evaluation}
The model achieved an R-squared score of 0.28 on the test set. This indicates that the model could use the inputs to explain 28\% of the variance in the output.

\subsubsection{Behavioural Investigation}
\begin{figure}[htbp]
  \centering
  \begin{subfigure}[b]{1\textwidth}
    \frame{\includegraphics[width=\textwidth]{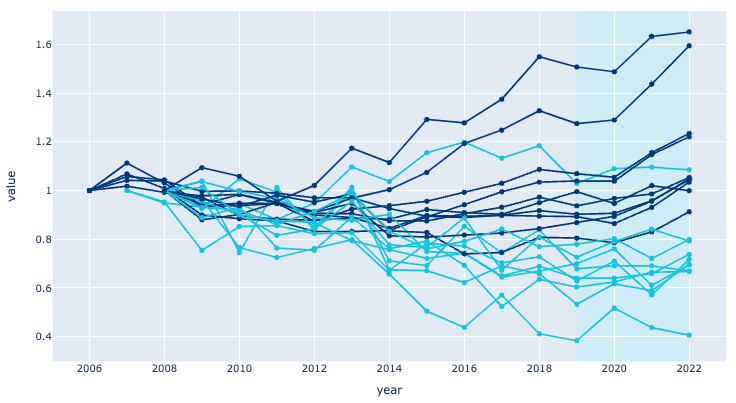}}
    \caption{Price Index}
    \vspace{2ex}
  \end{subfigure}
  
  \begin{subfigure}[b]{0.47\textwidth}
    \frame{\includegraphics[width=\textwidth]{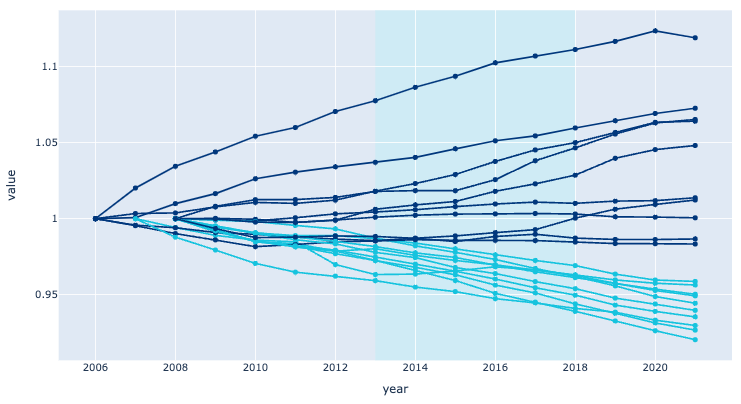}}
    \caption{Migrations}
    \vspace{2ex}
  \end{subfigure}
  \hfill
  \begin{subfigure}[b]{0.47\textwidth}
    \frame{\includegraphics[width=\textwidth]{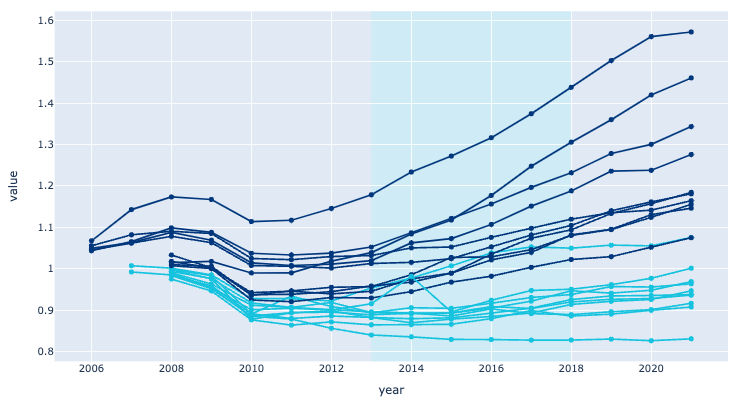}}
    \caption{Taxable Income}
    \vspace{2ex}
  \end{subfigure}
  
  \begin{subfigure}[b]{0.47\textwidth}
    \frame{\includegraphics[width=\textwidth]{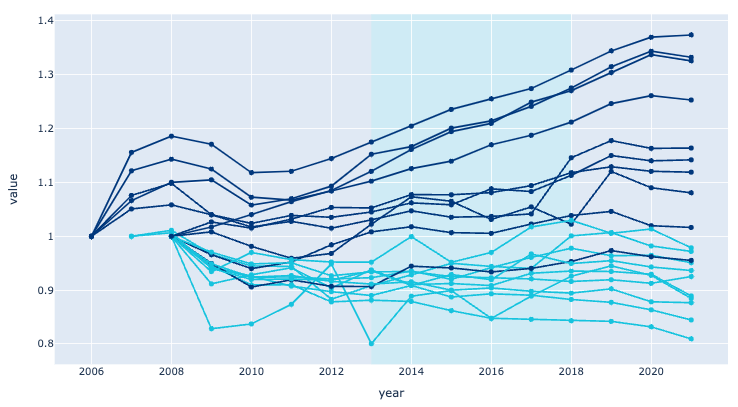}}
    \caption{Total Tax}
    \vspace{2ex}
  \end{subfigure}
  \hfill
  \begin{subfigure}[b]{0.47\textwidth}
    \frame{\includegraphics[width=\textwidth]{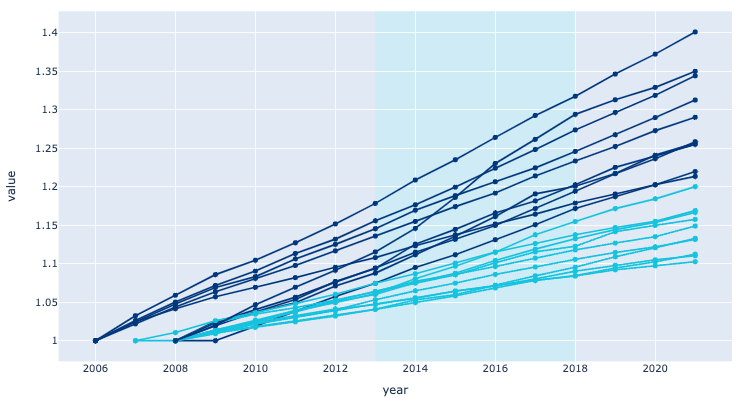}}
    \caption{Dwellings}
    \vspace{2ex}
  \end{subfigure}
  
  \caption{Difference in variables for outputs in the top decile and bottom decile}
  \label{fig:model_decile_comparison}
\end{figure}

To better understand the model's behaviour, we observed the input variation for 10 samples each from the top and bottom deciles of the model outputs. Figure~\ref{fig:model_decile_comparison} illustrates the model outputs, with darker blue representing outputs from the top decile. We see that the model's predicted top decile significantly outperforms the bottom. It's evident that the model weighs factors in line with the conclusions from Chapter 3.

\newpage
\section{Conclusion} \label{conclusion}

This paper investigated the use of alternative data variables and machine learning models to predict the performance of real estate markets in Japan. We first constructed a house price index at the municipality level using a comprehensive dataset of real estate transactions from 2005 to 2022. This allowed us to quantify the price appreciation or depreciation in each locality over time.

We then explored several socio-economic and demographic factors as potential predictors of real estate returns. Linear regression models showed that factors such as population growth, income levels and the number of newly constructed dwellings had statistically significant relationships with future price appreciation, although the explanatory power of each individual factor was limited.

To further test the predictive value of these alternative data variables, we devised a simple long-short investment strategy. Municipalities were ranked based on the past growth in each factor, and hypothetical portfolios were constructed by going long the top decile and short the bottom decile. The back-test results confirmed that in most cases, the factor-based portfolios outperformed a random investment strategy, providing some evidence of their predictive power.

Finally, we employed state-of-the-art Transformer models to predict the future price trajectory of each municipality based on the temporal evolution of the alternative data variables. The model was able to learn non-linear relationships and provide meaningful predictions, with the potential to improve further with the addition of new factors.

In conclusion, this paper demonstrates that alternative data variables do contain valuable information for predicting real estate market performance. Machine learning models can effectively harness this information to make meaningful predictions. However, more work is needed to refine the data sources, expand the factor set, and improve the model. With further research and development, these techniques can potentially provide a significant competitive advantage to investors in the real estate market.

\newpage

\bibliographystyle{ieeetr}
\bibliography{myreferences}

\end{document}